\documentclass[a4paper,fleqn,usenatbib]{mnras}

\usepackage{newtxtext,newtxmath}

\usepackage[T1]{fontenc}
\usepackage{ae,aecompl}

\usepackage{graphicx}	
\graphicspath{ {./figures/} }
\usepackage{amsmath}	

\usepackage{footnote}
\usepackage{amsmath}
\newcommand{\angstrom}{\text{\normalfont\AA}}

\title[Mapping Local ISM With DIBs]{Mapping Local Interstellar Medium With Diffuse Interstellar Bands}

\author[M. Piecka and E. Paunzen]{
Martin Piecka,$^{1}$\thanks{E-mail: 408988@mail.muni.cz}
and Ernst Paunzen$^{1}$
\\
$^{1}$Department of Theoretical Physics and Astrophysics, Masaryk University, 
Kotl\'a\v{r}sk\'a 2, CZ-611\,37 Brno, Czech Republic}

\date{Accepted XXX. Received YYY; in original form ZZZ}

\pubyear{2020}

\begin{document}
\label{firstpage}
\pagerange{\pageref{firstpage}--\pageref{lastpage}}
\maketitle

\begin{abstract}
With the use of the data from archives, we studied the correlations between the equivalent widths of four diffuse interstellar bands (4430$\angstrom$, 5780$\angstrom$, 5797$\angstrom$, 6284$\angstrom$) and properties of the target stars (colour excess values, distances and Galactic coordinates). Many different plots of the diffuse interstellar bands and their maps were produced and further analysed. There appears to be a structure in the plot of equivalent widths of 5780$\angstrom$ DIB (and 6284$\angstrom$ DIB) against the Galactic $x$-coordinate. The structure is well defined below $\sim150$ m$\angstrom$ and within $|x|<250$ pc, peaking around $x=170$ pc. We argue that the origin of this structure is not a statistical fluctuation. Splitting the data in the Galactic longitude into several subregions improves or lowers the well known linear relation between the equivalent widths and the colour excess, which was expected. However, some of the lines of sight display drastically different behaviour. The region within $150^\circ<l<200^\circ$ shows scatter in the correlation plots with the colour excess for all of the four bands with correlation coefficients $\textrm{R}<0.58$. We suspect that the variation of physical conditions in the nearby molecular clouds could be responsible. Finally, the area $250^\circ<l<300^\circ$ displays (from the statistical point of view) significantly lower values of equivalent widths than the other regions -- this tells us that there is either a significant underabundance of carriers (when compared with the other regions) or that this has to be a result of an observational bias.
\end{abstract}

\begin{keywords}
dust, extinction -- ISM: lines and bands -- ISM: molecules -- ISM: structure
\end{keywords}



\section{Introduction}

Diffuse interstellar bands (DIBs) are absorption spectral features, usually observed in the lines of sight towards hot stars (but also seen in spectra of other astronomical objects). To this date, several hundreds of different DIBs have been observed \citep{2008ApJ...680.1256H}, most of which are present in the visible part of the spectrum. Their discovery was made by \citet{1922LicOB..10..146H} while studying the sodium D lines, although the proper study of these features began in 1930s. The interstellar nature of the DIBs was discovered by looking at the spectra of binary stars \citep{1936ApJ....83..126M} where the position of the DIBs does not significantly change at the relevant timescales. Furthermore, it was also found that the DIBs are correlated with the colour excess $E(B-V)$ of the observed stars \citep{1936ApJ....83..126M,Snow77} which points to the fact that they are somehow related to the interstellar dust. The next step here would be the identification of the carrier of these bands. However, most of the ideas for carriers have been rejected. Atoms and diatomic molecules simply cannot fit the structure of the DIBs, while the studies of the polarisation at those wavelengths \citep{2007A&A...465..899C} show that the carriers are most likely not part of the dust population which is responsible for most of the polarisation of the starlight.

After almost 100 years, our knowledge of these mysterious interstellar bands has slightly improved. It was discovered that there are probably several families of the carriers \citep{1987ApJ...312..860K} and that some sort of structure can be identified in the profile of several individual bands \citep{1995MNRAS.277L..41S}. Since the discovery of fullerenes \citep{1985Natur.318..162K}, it was theorised that these molecules are going to be abundant in space and maybe also responsible for the formation of narrow absorption bands in the visible part of the spectrum. Although their presence in the universe was detected several years ago \citep{2010Sci...329.1180C}, it was only recently that the improvements in the mass-spectrometry allowed for an identification of at least two (and possibly three more) of the DIBs in the spectrum of the buckminsterfullerene C$_{60}^{+}$ \citep{2015Natur.523..322C} which is supported by space observation \citep{2019ApJ...875L..28C}.

Our work is focused on the usage of available data and spectra in the databases which can be used for mapping the DIBs in the near vicinity of the Solar System, and possibly beyond. It is possible that a comparison of these maps with other maps of the interstellar medium (ISM) can tell us a bit more about the nature of the carriers of the DIBs. Furthermore, these maps could be used as another diagnostic tool for the ISM, once these carriers are discovered. In the following sections, we will discuss the data sets used in this work which were retrieved from archives and the results which can be obtained from the plots of the equivalent widths (EWs) of the DIBs against different Galactic coordinates.

\begin{table*}
\caption{Data sets and the related errors of the measurements of the EWs. Third and fourth display the referenced spectral resolution and the number of measured lines of sight, respectively. We labelled a field N/A if no value is present in the respective work.}    
\label{table:1}   
\centering                          
\begin{tabular}{c c c c c c c c}         
\hline\hline      
 & Reference & R & LoS & Err (4430$\angstrom$) & Err (5780$\angstrom$) & Err (5797$\angstrom$) & Err (6284$\angstrom$) \\    
\hline              
    C13 & \citet{Chen13} & 22500 & 219 & N/A & N/A & N/A & 8 \% \\     
    G88 & \citet{Guarinos88} & N/A & 753 & 11 \% & 12 \% & 20 \% & 14 \% \\
    P13 & \citet{Puspitarini13} & 48000 & 129 & N/A & 35 \% & $>50$ \% & 50 \% \\
    R12 & \citet{Raimond12} & 48000 & 150 & N/A & 30 \% & N/A & 15 \% \\
    S77 & \citet{Snow77} & N/A & 2798 & N/A & N/A & N/A & N/A \\
    X11 & \citet{Xiang11} & N/A & 797 & N/A & 3 \% & 6 \% & 9 \% \\
\hline                                
\end{tabular}
\end{table*}

To this date, there have been only several attempts to map the DIBs on the global scale -- we will quickly mention some of the latest works. \citet{2016A&A...585A..12B} created a map of the DIBs based on observations of 670 nearby hot stars. On larger scale, \citet{2015ApJ...798...35Z} used the data from APOGEE and analysed over $60\,000$ lines of sight with the DIB at $1.527$ $\mu$m which resulted into good-quality maps covering about two thirds of the whole Galactic plane. Finally, \citet{2015MNRAS.452.3629L} analysed SDSS data and studied the lines of sight towards a significant number of stars ($\sim250\,000$), most of which are of cool spectral types. Although their map is very detailed, it is limited only to high Galactic latitudes (mostly $|b|>30 ^\circ$). On the other hand, this part of the sky complements very well the map from \citet{2015ApJ...798...35Z}.

Regarding the globular clusters, tracing DIBs is mostly used for studying the ISM between us and the targets. One of the earliest works was done by \citet{2009MNRAS.399..195V} who looked at the lines of sight towards $\omega$ Centauri cluster and studied the structure of the diffuse interstellar medium. Afterwards, more works follow -- for example, \citet{2015MmSAI..86..527M} studied the relation between the DIBs and the reddening towards the M4 cluster. \citet{2016MNRAS.463.2653D} also studied the relation between extinction and the DIBs -- they observed the DIB at 8620 $\angstrom$ towards Westerlund 1 and derived an empirical relation between the DIB and the extinction in the near infrared part of the spectrum. More recently, \citet{2017A&A...607A.133W} showed by observing the NGC 6397 cluster using VLT/MUSE that we now have instruments capable of studying the variations of the strength of the DIBs on quite small scales.

On the other hand, stellar associations and clusters can be used to study DIBs themselves and their carriers. \citet{2016ApJ...821...42H} observed spectra of stars in the Cygnus OB2 association and studied the relation between the carriers of the near infrared DIBs and the C$_2$ molecule. They found that these carriers and the C$_2$ molecule are not correlated which most likely means that the carriers are located, for the most part, in the diffuse part of the ISM.

\section{Archival Data}

The critical part of the problem when studying the DIBs is the lack of a significant number of publicly available observations. In the case of studying global properties of the DIBs, having a homogeneous sample of data (in terms of lines of sight) is quite important as well. Since most of these bands are very narrow (FWHM is typically below 1 $\angstrom$, although most of the stronger DIBs are somewhat wider) and some of them have low intensity, one needs a good spectral resolution and high signal to noise ratio -- we find the spectra with $\textrm{R}>15\,000$ and $\textrm{S/N}>100$ to be the best for studying the DIBs, although it is possible to use data with lower quality, as well. However, this will have a small effect on the error estimates of the EW measurements. Moreover, some DIBs may be difficult (or impossible) to measure in low-quality spectra.

Unfortunately, the data present in archives usually contain information about only a few of the strongest DIBs, namely 4430$\angstrom$, 5780$\angstrom$, 5797$\angstrom$ and 6284$\angstrom$. For the purposes of our work, we have chosen the data sets from the VizieR database which are listed in Table \ref{table:1}. The displayed uncertainties of the values from EWs were calculated as the median of all values for a given DIB. It should be noted that not all of the data sets contain the information about all four chosen bands but all of them contain the information about reddening towards given lines of sight.

From now on, the referenced data sets will be abbreviated according to the Table \ref{table:1}. There are several facts about these data sets which need to be pointed out. Firstly, S77 contains no information about the uncertainties of determination of the EWs. Since this set contains almost half of the combined data, we adopted the median values of uncertainties of the EW measurements in the other data sets. Secondly, there are two sets of values of the 6284$\angstrom$ EWs in C13 -- we note that for this work we used the EW values after correction. Furthermore, the target stars in C13 are objects in the Baade's Window and are of cool spectral types, unlike most of the other target stars. For this reason, we should be careful when comparing these data with the ones from other data sets. Finally, there is an outlier point in S77 in the 5797$\angstrom$ DIB data which is probably just a bad measurement. We have decided to remove this value and replace it with the one from X11 (for the same target star).

Since we also aim to explore the relation between DIBs and the rectangular Galactic coordinates, we need to know the distances/parallaxes assigned to the target stars. These can be taken from SIMBAD using an automatic procedure which uses the names of the target stars that are listed in the VizieR database. However, not all of the stars have had their parallax measured, so the amount of data used for this part of the work is somewhat reduced. For example, data from C13 are completely absent in SIMBAD and we were unable to obtain specific values of the distances in any other way (althoug in this case the knowledge of the distance would give us little to no additional information). Furthermore, it should be pointed out that the parallaxes obtained from SIMBAD were measured by both, Hipparcos and Gaia (Gaia-DR2) \citep{1997A&A...323L..49P,2018A&A...616A...1G} space observatories. Since presently Hipparcos data form only an extremely small fraction of the whole set used in this work, we do not have to worry as much about the significant error of measurements present in the data provided by Hipparcos. It is also worth mentioning that most of the target stars are located within 1 kpc from the Sun -- the ISM mapped using the data can be, therefore, considered as "local".

In order to investigate the astrophysical properties of our targets in a colour-magnitude diagram (CMD), 
we use the homogeneous Gaia-DR2 photometry. We have excluded stars with no parallax and photometric measurements.
Furthermore, we restricted the analysis to stars with parallax errors less than 20~\% in order to avoid
a significant bias to the statistics of the sample \citep{Francis2014}. This left us with 531 objects which we used to investigate biases in the CMD
(the rest of the analysis in the following sections was based on the complete sample).
 
The transformation of the reddening values was performed using the following relations
\begin{equation}
A_V = 1.1 \, A_G = 3.1 \, E(B - V) = 2.1 \, E(BP - RP).
\end{equation}

\begin{figure}
\begin{center}
\includegraphics[width=80mm, clip]{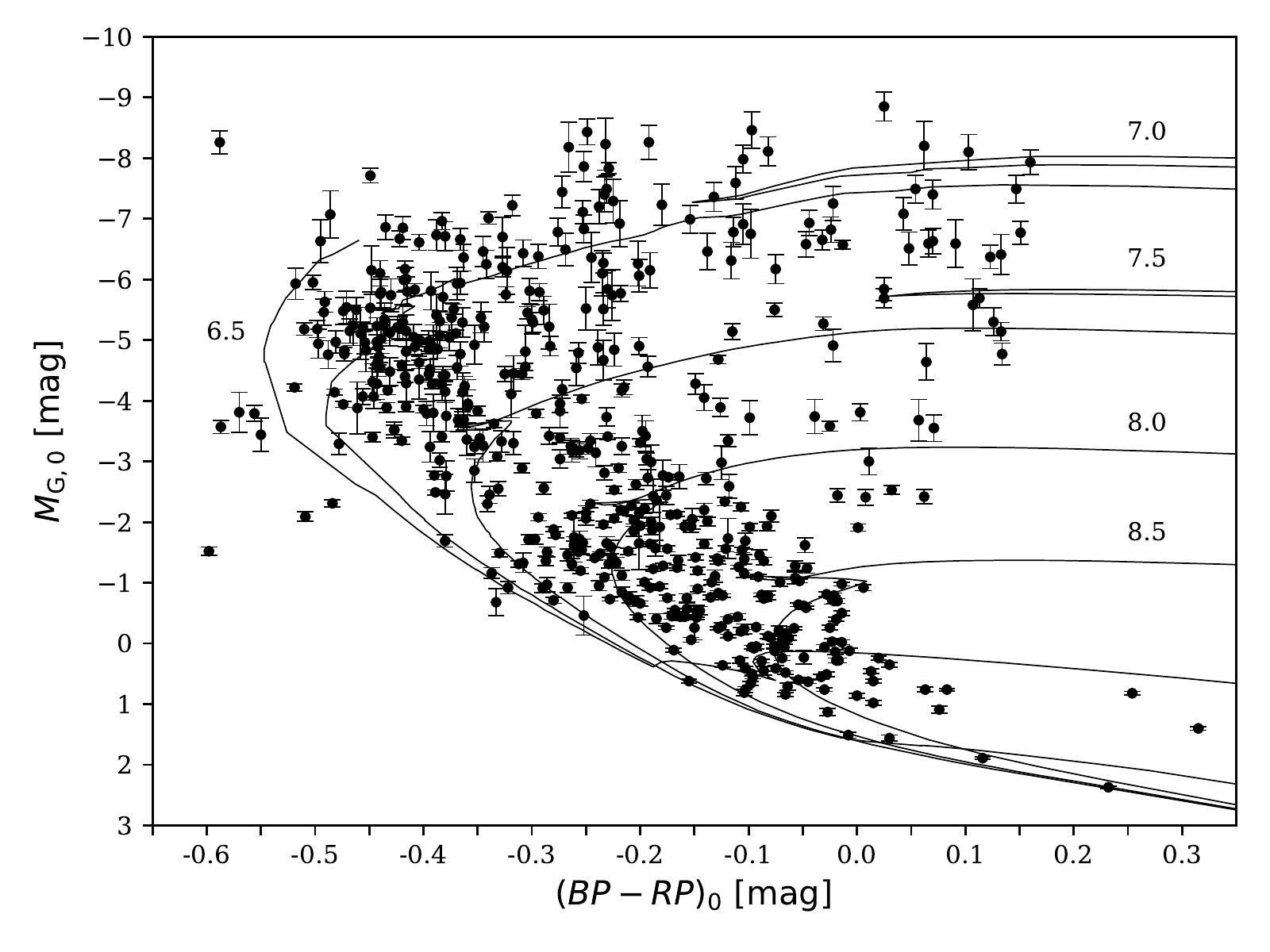}
\caption{The $(BP - RP)_0$ versus $M_{\mathrm{G,0}}$ diagram of our targets together with the 
PARSEC isochrones (listed are the logarithmic ages) for a solar metallicity of Z\,=\,0.02.}
\label{HRD} 
\end{center} 
\end{figure}

The absolute magnitudes were calculated using the distances from \citet{Bailer2018}.
In Fig. \ref{HRD}, we present the CMD of our targets together with the 
PARSEC isochrones \citep{Bressan2012} for a solar metallicity of Z\,=\,0.02. We favour this value because
it is consistent with the recent results of Helioseismology \citep{Vagnozzi2019}. A smaller value, as also suggested by
\citet{Bressan2012}, shifts the main sequence slightly to the blue. The overall estimated 
reddening values seem very reliable because only
eight objects (BD+40 4220, HD 1810, HD 30112, HD 36960, HD 175803, HD 181963, HD 191639, and HD 202349) are located 
(within 3\,$\sigma$) below the zero age main sequence. The absorption $A_G$ for HD 30112 is 5.6\,mag. A wrong extinction value from
the literature would explain its location in this diagram. For all other objects, the absorption is only small.

It is worth mentioning that there are some differences between inverting parallaxes to obtain distances from Gaia-DR2 and the distances
calculated by an alternative procedure from \citet{Bailer2018} (which is also using data from Gaia-DR2). Although these differences are mostly minor (especially for distances below 2 kpc the correlation is almost $1.0$), there are still some outliers and so we have decided to use this catalogue since it promises more accurate distances. Inversion of parallax was used only in the case when \citet{Bailer2018} do not give the distance for a star.

If we investigate the distribution of stars in Fig. \ref{HRD}, we find that one sample is concentrated for ages below
40\,Myr and one above 100\,Myr. This characterizes the stars within OB stellar associations and the Galactic field, respectively. 
Furthermore, there is a significant fraction of stars which have already left the zero age main sequence. These objects
may have significant stellar winds which have been connected to the formation or destruction of the carriers of DIBs \citep{2017MNRAS.470.2835L}.

\section{Maps Of The Local ISM}

We define the rectangular Galactic coordinates as per usual -- we choose the $x$-axis to be the direction towards the Galactic centre, the $y$-axis is perpendicular to $x$-axis and is oriented towards the direction of Galactic rotation and the $z$-axis is perpendicular to the $x$-$y$ plane and orientated towards the Galactic north pole. This coordinate system $(x,y,z)$ can be easily transformed into the right-handed spherical Galactic coordinates $(l,b,r)$. The coordinate $r$ represents the distance between us and the observed object, Galactic longitude $l$ represents the angle between the Galactic centre and the observed object and the Galactic latitude $b$ represents the angle between the $x$-$y$ plane and the observed object.

\begin{figure}
  \resizebox{\hsize}{!}{\includegraphics{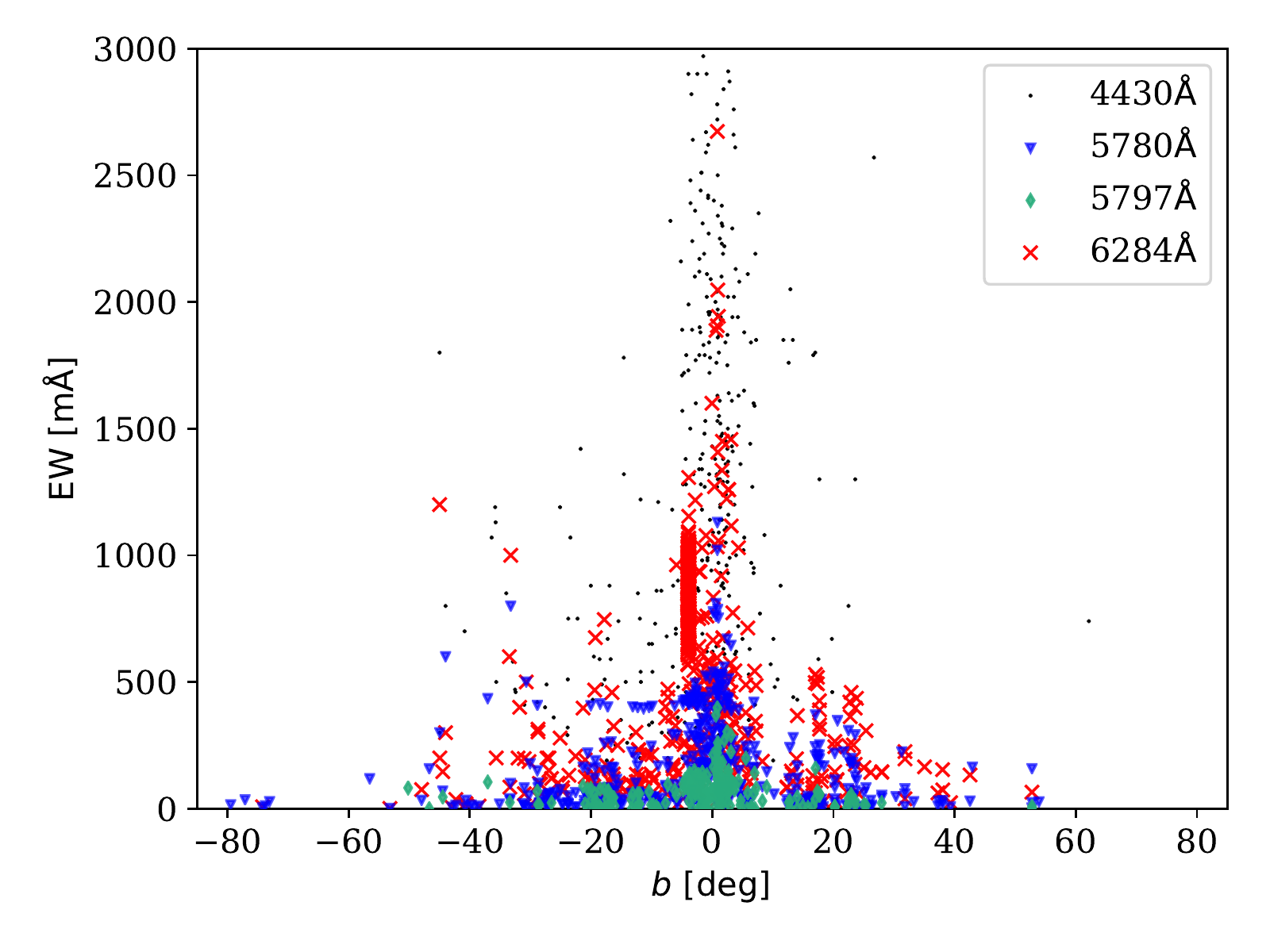}}
  \caption{Distribution of the strength of the DIBs in Galactic latitude $b$. Black dots represent 4430$\angstrom$, blue triangles 5780$\angstrom$, teal diamonds 5797$\angstrom$ and red crosses 6284$\angstrom$.}
  \label{fig01}
\end{figure}

\begin{figure}
  \resizebox{\hsize}{!}{\includegraphics{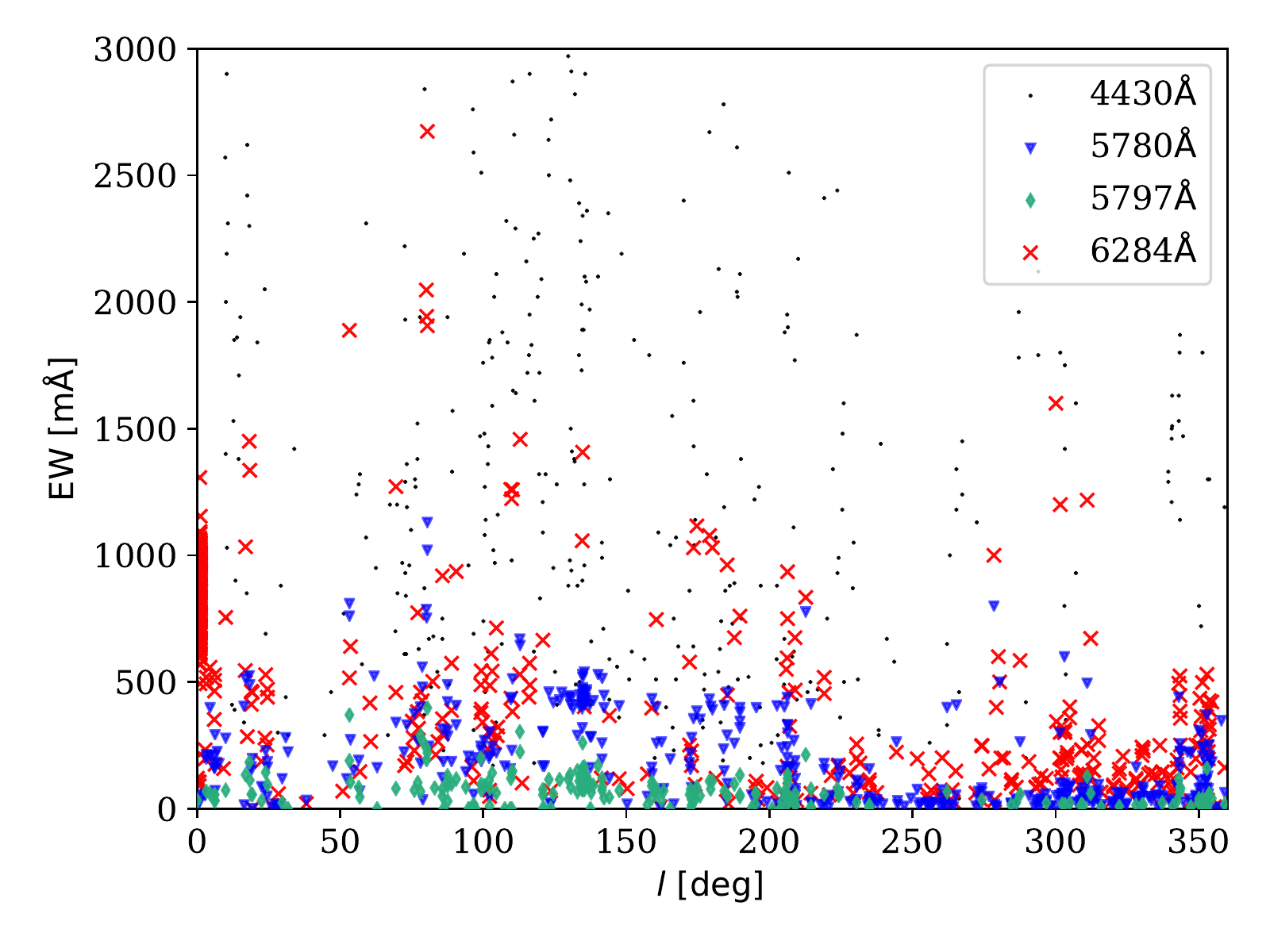}}
  \caption{Distribution of the strength of the DIBs in Galactic longitude $l$. Black dots represent 4430$\angstrom$, blue triangles 5780$\angstrom$, teal diamonds 5797$\angstrom$ and red crosses 6284$\angstrom$.}
  \label{fig03}
\end{figure}

Using the joint data from the VizieR database, we have created maps of the four DIBs at 4430 $\angstrom$, 5780 $\angstrom$, 5797 $\angstrom$ and 6284 $\angstrom$. Combined data of all four bands are display in Fig.~\ref{fig01} and Fig.~\ref{fig03}. The rest of the plots show mostly the $5780\angstrom$ band as a prototype of the results. The rest of the plots are shown in the appendix but the comments on these are presented in this section.

\begin{figure}
  \resizebox{\hsize}{!}{\includegraphics{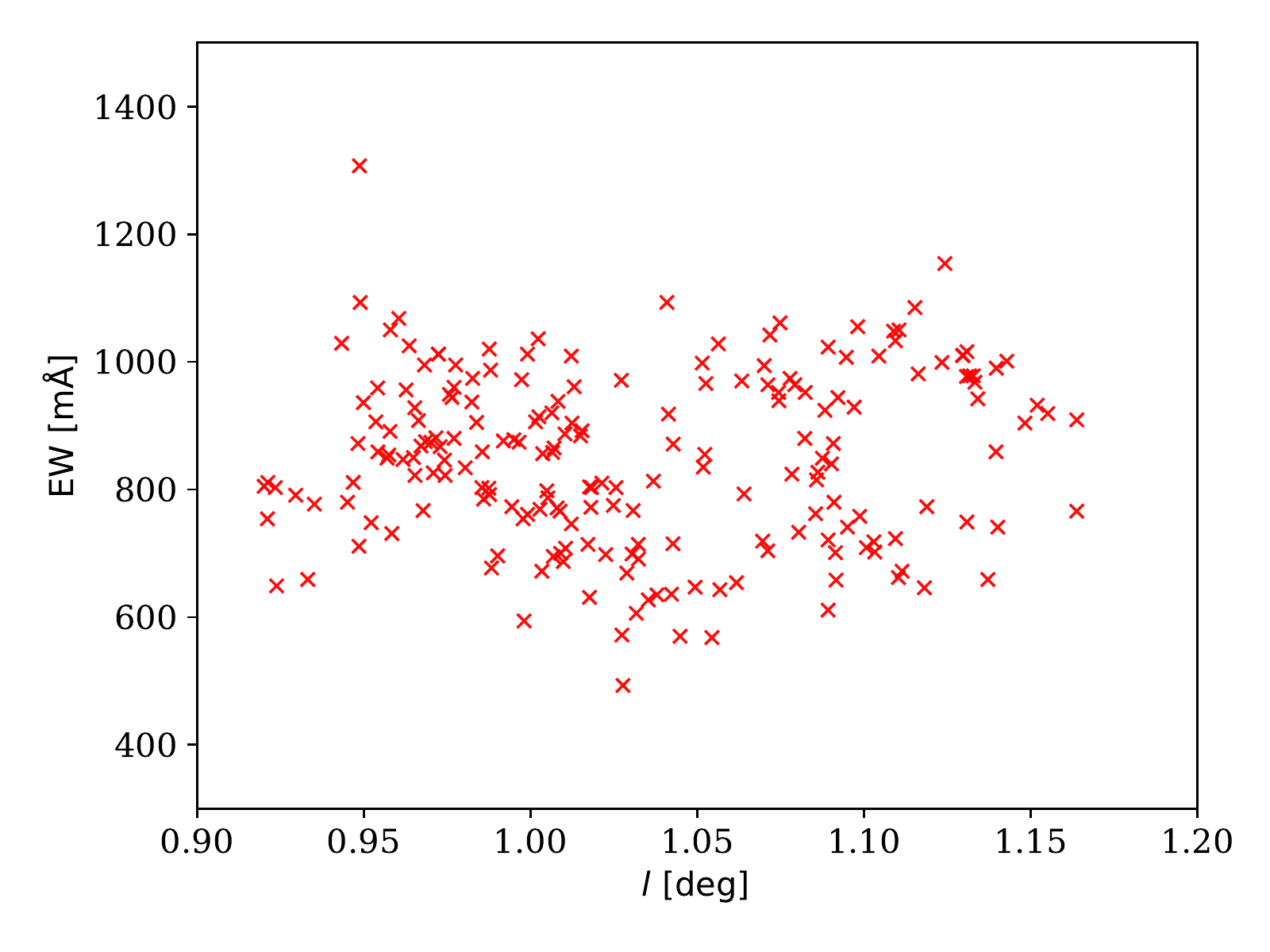}}
  \caption{Strength of the 6284$\angstrom$ DIB in the lines of sight towards the Baade's Window.}
  \label{fig04}
\end{figure}

\begin{figure}
  \resizebox{\hsize}{!}{\includegraphics{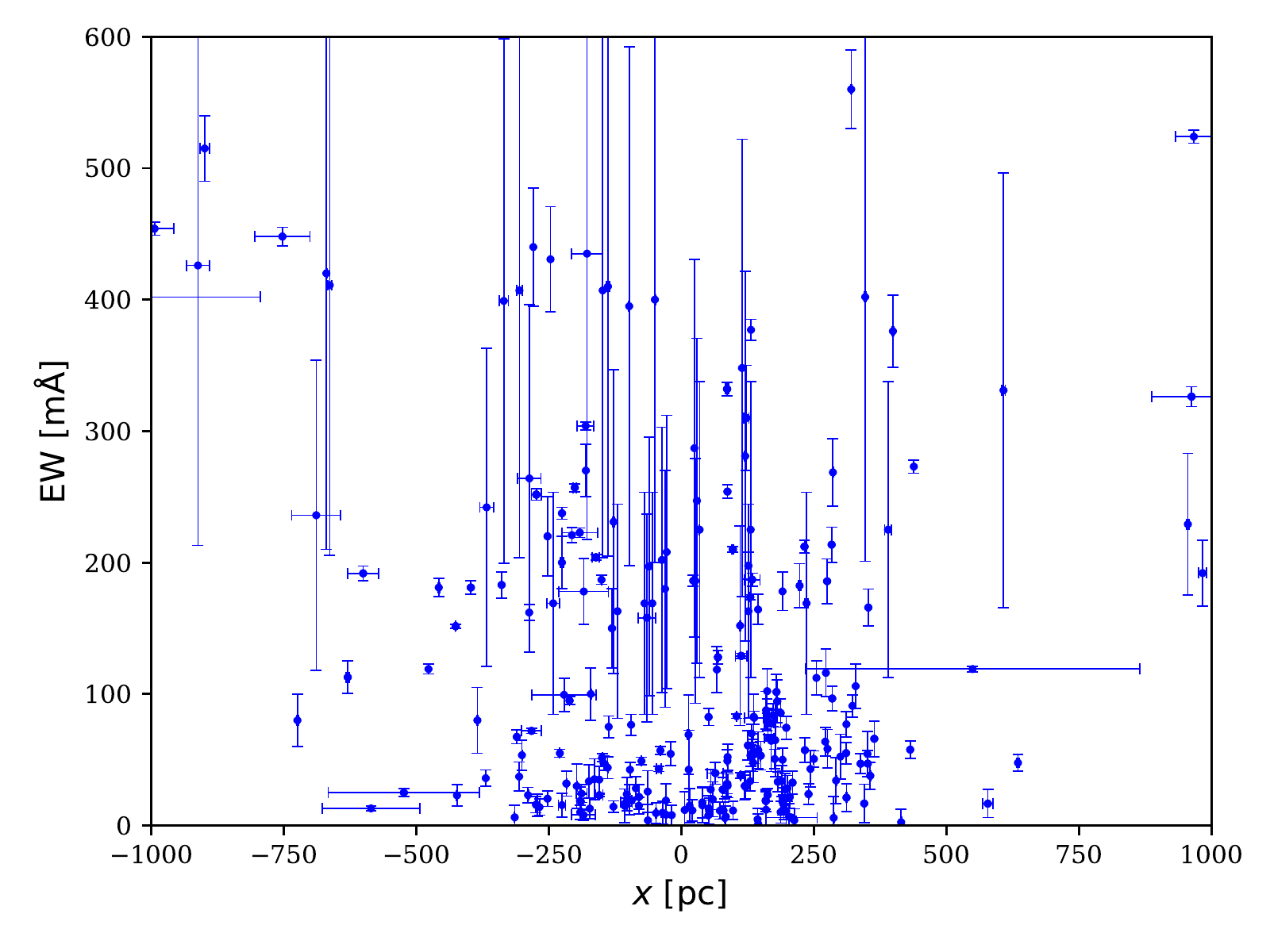}}
  \caption{Relationship between the strength of the 5780$\angstrom$ DIB and the Galactic $x$ coordinate of the target stars, with displayed error bars. The structure in the lower part of the plot seems to be well constrained within the uncertainties which means that it is most likely real and not a product of random distribution.}
  \label{fig06}
\end{figure}

\begin{figure}
  \resizebox{\hsize}{!}{\includegraphics{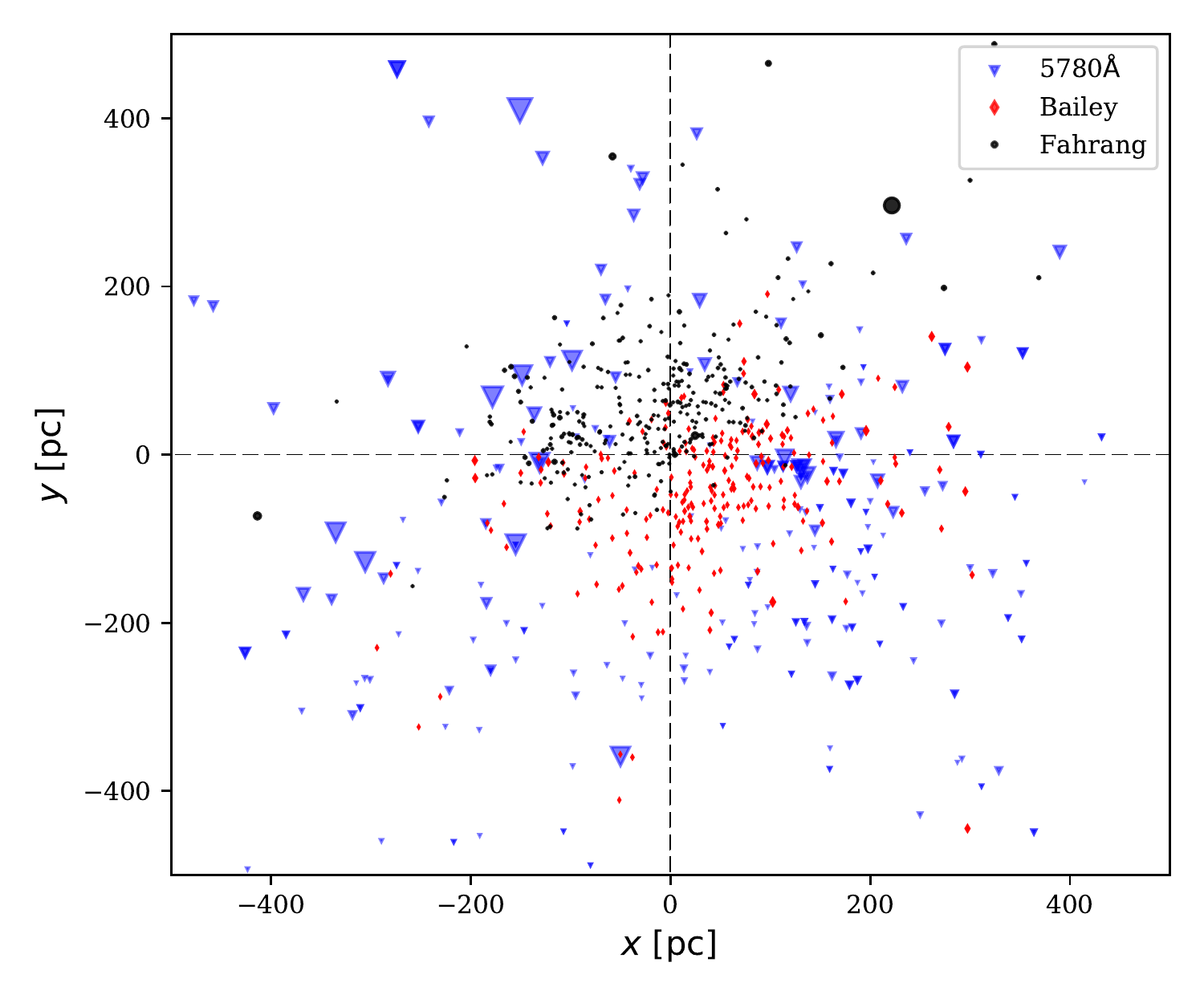}}
  \caption{Distribution of the target stars of the used sample for the band at 5780$\angstrom$. Blue triangles show the data we present in this paper, black circles and red diamonds are based on the data taken from VizieR -- \citet{2019NatAs...3..922F} and \citet{2016A&A...585A..12B}, respectively. Size of a symbol refers to the strength of the DIB in the spectrum of a given star -- symbols are larger as the DIB gets stronger.}
  \label{fig08x}
\end{figure}

The EW plotted against the Galactic latitude can be seen in Fig.~\ref{fig01}. We can clearly see that the strength of the DIBs falls down as we look further up. The same can be seen for the Galactic coordinate $z$ which is displayed in Fig.~\ref{figZ}, although the uncertainties in the $z$ coordinate can be quite high and affect what we see. The values of the EWs seem to decrease up to about 500 pc which would correspond to the fact that the carriers of the DIBs are confined to the Galactic disk. However, it should be noted that this can also be the result of the bias caused by the lack of stars observed at $|z| >500$ pc. Although it is, in principle, possible to determine the shape of the structure seen in Fig.~\ref{fig01}, we advise against it due to the fact that there are many points without known uncertainties. It is also worth mentioning that for the 5780$\angstrom$ DIB there are data points ($\textrm{EW} \sim 400$~m$\angstrom$) in the latitude plot which seem to have shifted zero-point value. Further discussion of this issue is presented in the next section, after we have had a look at another important plot.

There seems to be a very complex behaviour in the plots of the Galactic longitude (Fig.~\ref{fig03}). Generally for all the DIBs, there seems to be a maximum somewhere between $l=0^\circ$ and $l=150^\circ$, while a minimum can be located at about $l=250^\circ$. However, there is much more going on if we focus on just one of the plots. There appear to be at least two local maxima ($l=0^\circ$, $l=80^\circ$) and at least two local minima \mbox{($l=50^\circ$, $l=250^\circ$)}. On the other hand, the region \mbox{$80^\circ < l < 250^\circ$} displays a behaviour which cannot be displayed in the simple terms of minima and maxima alone. We have chosen the $5780\angstrom$ band as a prototype since the behaviour for the $5797\angstrom$ and $6284\angstrom$ bands is quite similar. The data from C13 give us the opportunity to study the $6284\angstrom$ plot in more detail in a small area of the sky. As can be seen in Fig.~\ref{fig04}, there is an apparent structure (upper outline of the distribution of points) in the Galactic longitude plot even on the scales of $0.1^\circ$ -- it needs to be pointed out that the typical value of uncertainties in Fig.~\ref{fig04} is about 8~\%. The detailed structure of the last DIB at $4430\angstrom$ appears to significantly differ from the other ones. Unlike in the case of the other DIBs, we do not observe any striking variation in the apparent structure except of a wide minimum in the right half of the plot.

The existence of the mentioned structure is questionable. Since we expect the number of carriers to increase as we look farther away, there must be some (likely very complicated) relation between the apparent surface values of the distribution of the EWs and the distances towards the observed objects. This could be explored in detail by looking at different distances if more measurements of DIBs in different lines of sight at different distance were available in the archives.

Fig.~\ref{fig06} shows the plot of the equivalent width of the DIB at 5780~$\angstrom$ against the Galactic coordinate $x$ which represents the direction towards the Galactic centre (see Fig.~\ref{figX} for the other DIBs). In this direction, if we concentrate on the area below the horizontal line at $\textrm{EW}=150$ m$\angstrom$, we will find a structure that can be defined at $|x| <250$ pc -- this is apparent also in the plot for 6284$\angstrom$ band but not for the other two DIBs. This structure reaches a maximum at the distance $x=170$ pc. We see a scatter of points at values $\textrm{EW}>200$ m$\angstrom$. Before jumping to conclusions, we need to consider whether what we see can really be there. First, we have two sources of uncertainties which need to be taken into account. The error in the $x$ coordinate is taken to be the actual error in the distance towards the target star -- this is the upper limit. Next, since there are many measurements from S77, we need to estimate uncertainties for the EWs. For this, we take a look at the typical values mentioned in Table \ref{table:1} and take 50~\% as the uncertainty for all EW measurements from S77. Although this approach is somewhat arbitrary, it may be the only way of estimating the uncertainties for S77. The result is shown in Fig.~\ref{fig06} which displays only the points with errors equal or lower than 30~\% in the $x$ coordinate and $0.0<E(B-V)<4.0$ mag. If we focus on the structure we defined in this paragraph, we can see that the error bars allow its existence. There may still be some sort of bias due to the amount of observations made towards positive and negative direction of $x$, but we see a lot of measurements on both sides and conclude that this is unlikely, mostly due to what we see in Fig.~\ref{fig06}. However, we cannot dismiss the argument that a bias may still actually be there -- for example, the zero-point shift (discussed above) may play some role in this.

\begin{table*}
\caption{The correlation coefficients of linear fits of the data in different lines of sight. Asterisk is added for regions where the total number of data points is less than 10.}   
\label{table:2}
\centering
\begin{tabular}{c c c c c c} 
\hline\hline 
Region & Name & r (4430$\angstrom$) & r (5780$\angstrom$) & r (5797$\angstrom$) & r (6284$\angstrom$) \\ 
\hline  
                                            & Original (unseparated) data & 0.675 & 0.815 & 0.850 & 0.704 \\ 
    $-60^\circ < l < 40^\circ$ & Galactic-central region & {\color{red} 0.612} & {\color{red} 0.779} & {\color{green} 0.942} & {\color{green} 0.774} \\
    $40^\circ < l < 75^\circ$ & first-minimum region & {\color{green} 0.782} & {\color{green} 0.923} & {\color{green} 0.876}\rlap{*} & {\color{red} 0.653} \\
    $75^\circ < l < 125^\circ$ & first-peak region & {\color{red} 0.633} & {\color{green} 0.871} & {\color{green} 0.897} & {\color{green} 0.740} \\
    $125^\circ < l < 150^\circ$ & 5780$\angstrom$ double-trend region & {\color{green} 0.681} & {\color{red} 0.553} & {\color{red} 0.616} & {\color{green} 0.780}\rlap{*} \\
    $150^\circ < l < 200^\circ$ & scatter-field region & {\color{red} 0.577} & {\color{red} 0.338} & {\color{red} 0.299} & {\color{red} 0.553} \\
    $200^\circ < l < 250^\circ$ & unknown-trend region & {\color{green} 0.743} & {\color{red} 0.628} & {\color{red} 0.718} & {\color{green} 0.918} \\
    $250^\circ < l < 300^\circ$ & wide-minimum region & {\color{green} 0.746} & {\color{red} 0.745} & {\color{red} 0.477}\rlap{*} & {\color{green} 0.821} \\
\hline                        
\end{tabular}
\end{table*}

\begin{figure}
  \resizebox{\hsize}{!}{\includegraphics{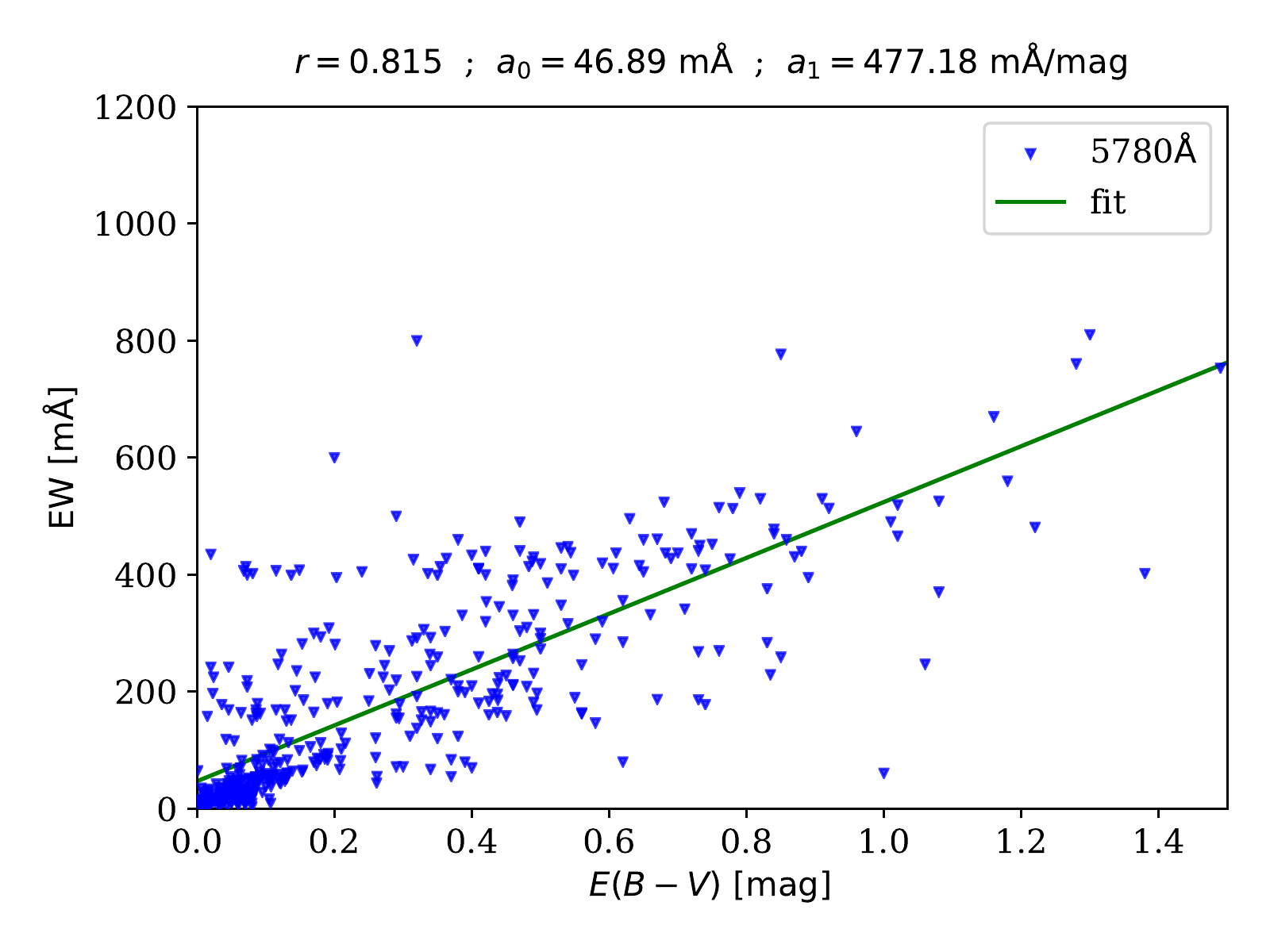}}
  \caption{The correlation between the equivalent width of the 5780$\angstrom$ and the colour excess $E(B-V)$.}
  \label{fig09}
\end{figure}

We have taken the same approach with the Galactic coordinate $y$, defined by the direction of Galactic rotation. Once again, we can see some sort of structure in the plot of the equivalent width of the DIB at the 5780$\angstrom$ DIB (Fig.~\ref{figY}) in the area $\textrm{EW}<150$ m$\angstrom$, \mbox{$-500$ pc $< y < 0$ pc}. This time it seems obvious that the result is biased -- the plot shows only a few measurements in the positive direction of $y$. In order to check this, we have created simple plots of the Galactic disk plane with the size of the points indicating the value of the EW -- this is displayed in Fig.~\ref{figXY}. Looking closely at this map, we see that there is a relative lack of lines of sight containing higher amounts of the species producing the 5780$\angstrom$ feature in the plane ahead of us compared with the plane behind us. This map can be directly compared with the one from \citet{2016A&A...585A..12B} where we see that both maps are quite complementary. Moreover, we can see in Fig.~\ref{fig08x} the comparison with the data from \citet{2016A&A...585A..12B} and the recent work by \citet{2019NatAs...3..922F} who studied mostly objects within 200 pc. Data from this work complete our picture of the map within 100 pc radius area where data from Table \ref{table:1} is lacking in the number of observations.

\section{Trends In The Reddening Correlations}

Let us now take a look at the correlations between the EWs and the colour excess $E(B-V)$. As can be seen in Fig.~\ref{fig09}, there appears to be an almost linear relation which was already established by \citet{1936ApJ....83..126M}. However, this trend is quite broad which means that the relationship between the carriers and the reddening is a bit more complex. With the use of the results shown in the previous section we attempted to search for more narrow trends in these plots. In this section, we will only look at the plots for the 5780$\angstrom$ DIB -- the rest of them are presented in the Appendix.

\begin{figure}
  \resizebox{\hsize}{!}{\includegraphics{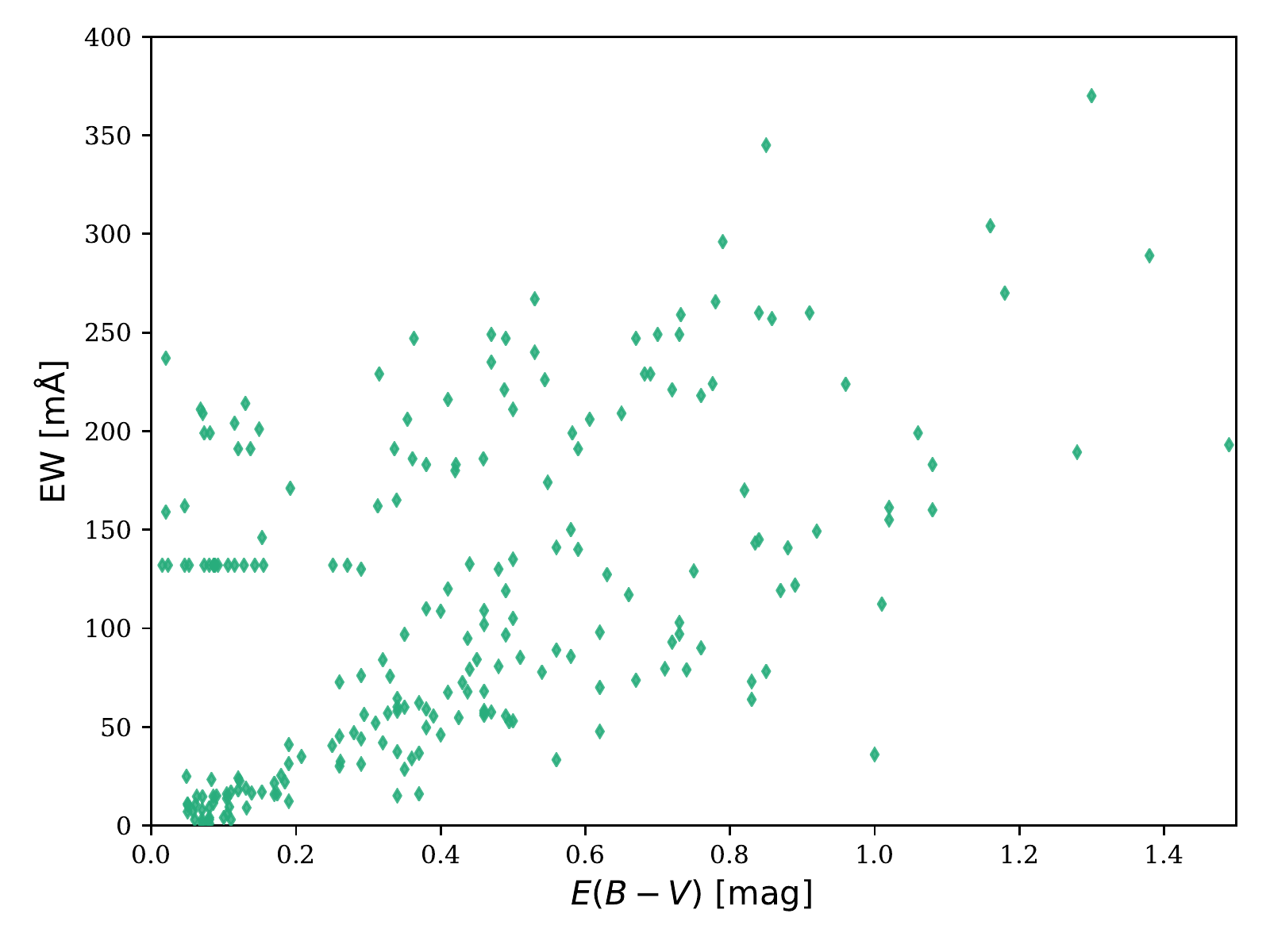}}
  \caption{The correlation between the equivalent width of the 5797$\angstrom$ DIB and the colour excess $E(B-V)$. Some points display an obvious shift of their zero-point value ($\sim 130$ m$\angstrom$), similar to the 5780$\angstrom$ DIB ($\sim 400$~m$\angstrom$). These data points correspond to the same objects for both DIBs, with EWs presented in S77.}
  \label{fig10}
\end{figure}

\begin{figure}
  \resizebox{\hsize}{!}{\includegraphics{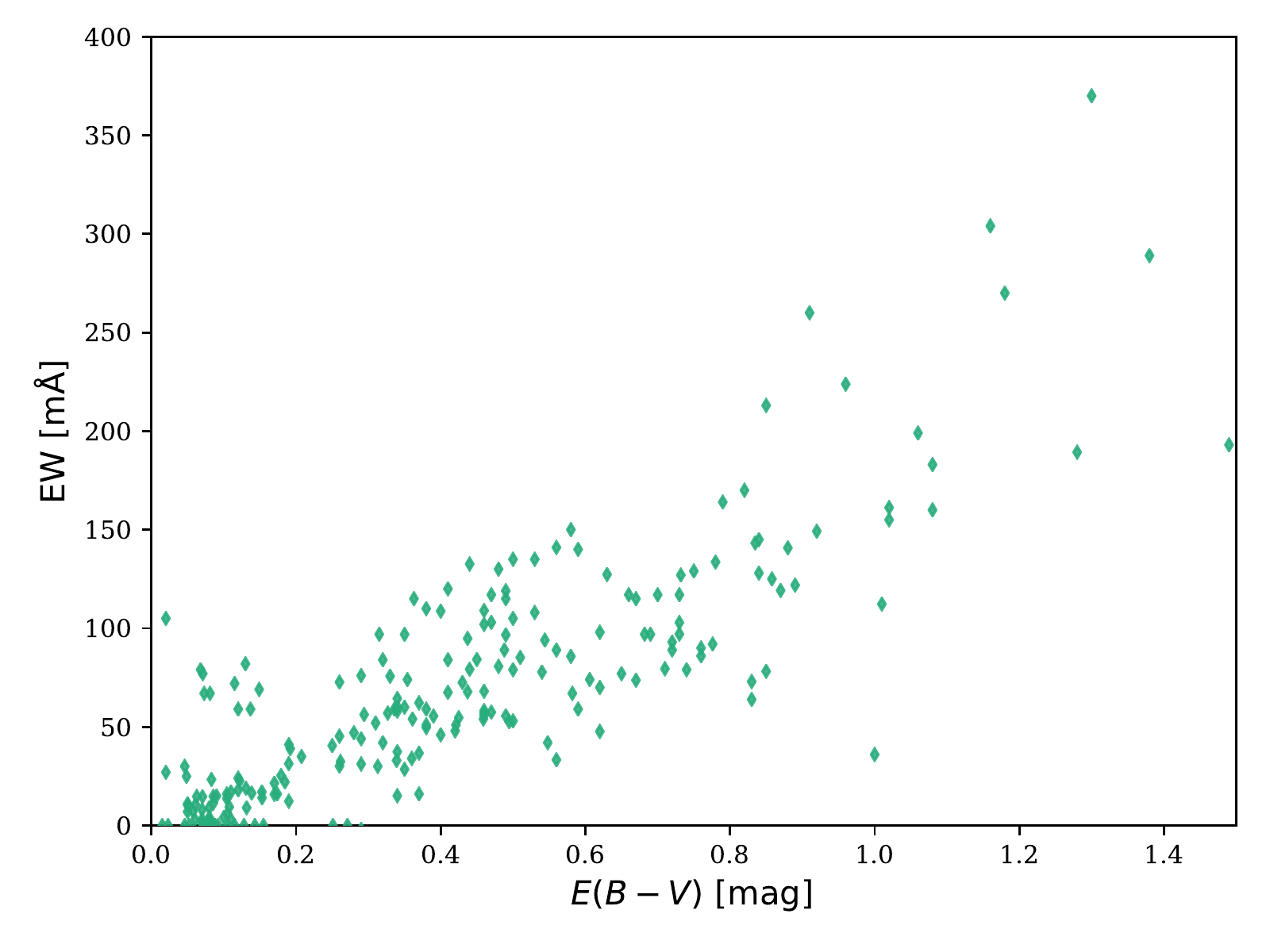}}
  \caption{The correlation between the equivalent width of the 5797$\angstrom$ DIB and the colour excess $E(B-V)$. Some of the points where shifted downwards by 132 m$\angstrom$ in order to correct for the zero-point shift present in the data.}
  \label{fig11}
\end{figure}

Before we start analysing the data, we need to take a closer look at the 5797$\angstrom$ DIB in Fig.~\ref{fig10}. It can be clearly seen that there are two different trends in the data. If we pay attention to the details of the plot we can see that there is a line of points at $\textrm{EW} \sim 140$ m$\angstrom$. This seems like a zero-point value part of the upper trend in the plot -- it appears that some of the data are shifted towards higher values, possibly by the same amount as the zero-point value part. For this reason, we have decided to shift the data points in the upper trend by this value downwards (Fig.~\ref{fig11}) which improves what we see in all of the plots that include the 5797$\angstrom$ DIB. We did this by looking at points above the line $\textrm{EW}_{5797}=154.3 \, E(B-V) + 73.0$ and shifting them by the value of the upper zero-point value (132 m$\angstrom$). Unfortunately, we do not know which points are part of this shifted trend and which are not, which means that such procedure will insert a possible bias in the data set. Furthermore, the data can only be shifted by separating the trends by a line which is chosen almost arbitrarily -- there is no reason to choose the parameters of the line that we used over slightly different values. We would also like to point at the 5780$\angstrom$ DIB where we also see a similar shift (Fig.~\ref{fig09}). However, in this case we cannot make any corrections due to the fact that we cannot precisely define the zero-point value. Moreover, the spread of points in Fig.~\ref{fig09} is somewhat larger than in Fig.~\ref{fig10}. Finally, it should be mentioned that this shift is exclusive to S77 (not all measurements display this behaviour) and cannot be seen in the other data sets.

We have separated the data into 7 different sets, described in Table \ref{table:2}, based on the details that can be seen in the plots of the Galactic longitude. In the table, the values highlighted in green show improvement (in terms of correlation coefficients) from the original correlation, the red colour indicates lower value for the correlation coefficient and asterisk denotes regions where the total number of data points is less than 10. The names of the regions defined by these different values of longitude are related to what we see in the plots. For example, the first minimum region is related to the first obvious minimum in the longitudinal plot. Also, for each of the plots we fitted a linear function and determined the correlation coefficient for the given region. In principle, we should expect each region to have a higher value of this coefficient than we get if we do not distinguish between the longitudinal regions. However, it must be emphasised that there is an error present in the measurement of the EWs and that we do not have the amount of measurements required for this "detrending" method to work perfectly. On the other hand, we can expect that some of the regions should display at least slight improvement, even with the data we use.

By looking at the results of the separation of regions by longitude and comparing with the original correlation coefficients, we can safely say that there really are many different trends in the original (unseparated) reddening plots. The increase in the value of the correlation coefficient can be best seen for the 5797$\angstrom$ and 6284$\angstrom$ DIBs (in some lines of sight). On the other hand, there is only a small improvement for the 4430$\angstrom$ DIB in terms of the coefficients. Finally, the 5780$\angstrom$ DIB seems to be well correlated with the reddening only in several lines of sight and possibly displays a very complex behaviour in the direction towards the region $125^\circ < l < 150^\circ$.

\begin{figure}
  \resizebox{\hsize}{!}{\includegraphics{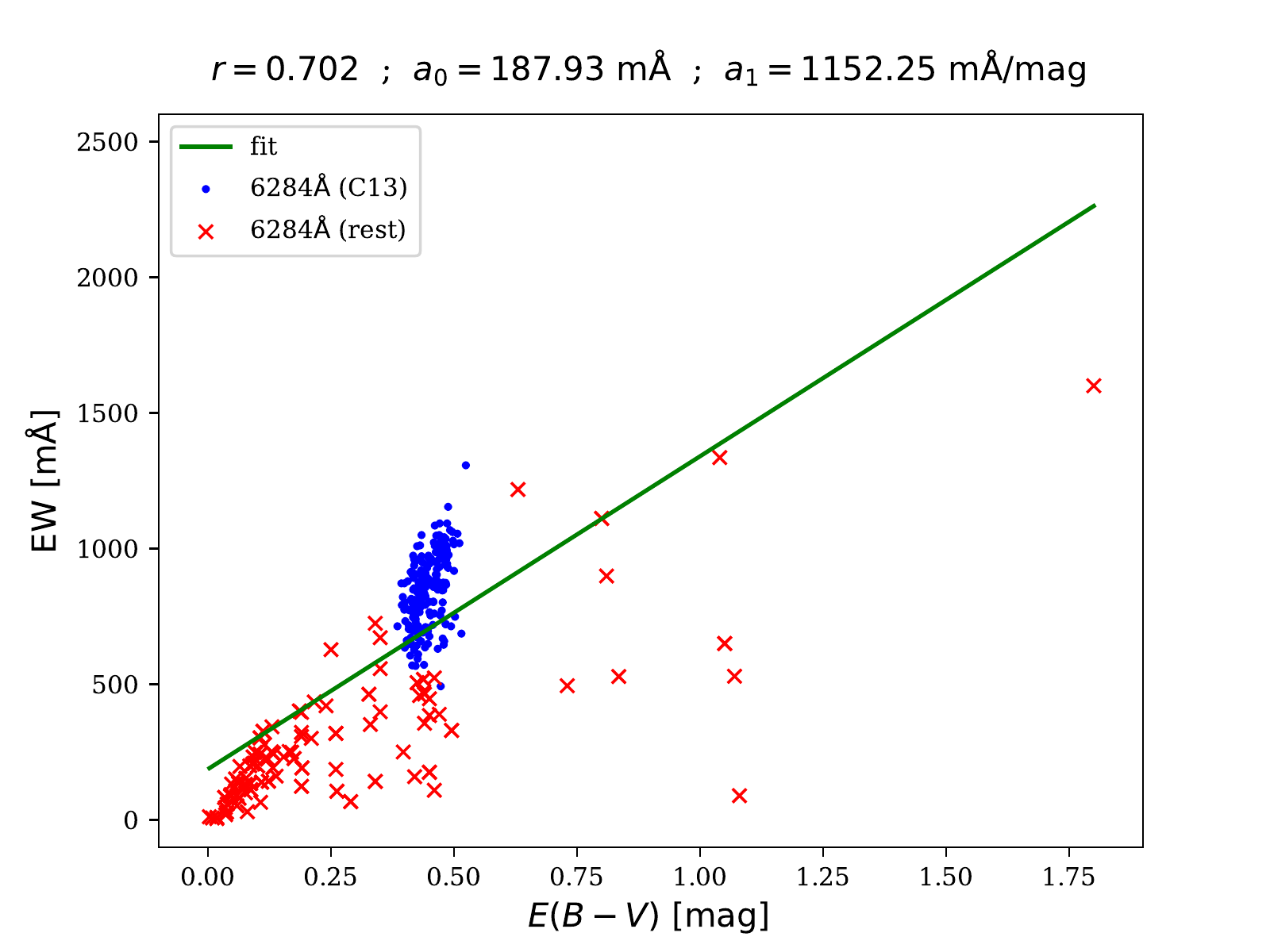}}
  \caption{The correlation between the colour excess $E(B-V)$ and the strength of the 6284$\angstrom$ DIB in the Galactic-central region. The second trend (blue crosses) corresponds to the stars in the Baade's Window.}
  \label{fig13}
\end{figure}

Let us discuss these results in detail. To start with, the Galactic-central region seems to be the dominant part of the original data since the correlation line does not significantly differ from the original one for all of the DIBs (see Fig.~\ref{figtest0} and Fig.~\ref{figtest1}). This is an important result because if there are other trends in the unseparated data, observing mostly in these directions will produce a bias. Moreover, we can also take a closer look at the 6284$\angstrom$ DIB (Fig.~\ref{fig13}) and see the behaviour of the data from C13 -- these clearly follow a very different trend than the rest of the data in this plot (if we had separated them) but also perfectly fit to the line in the plot which means that the correlation would not change much if we had not used the data from C13.

The first-minimum region improves the correlation for all of the DIBs but the one at 6284 $\angstrom$ (Fig.~\ref{figtest2}). However, not much else can be said about this region since there are only a few measurements in this region for all of the DIBs we work with.

The first-peak region contains a slightly larger amount of data. We can see some improvements from the unseparated data in terms of the coefficients. However, the shape of the correlation must be considered, as well. Although the correlation value is smaller for the 4430$\angstrom$ DIB, the correlation can be clearly seen in the plot and it appears to be quite consistent, although broad (Fig.~\ref{figtest3}, upper left panel). On the other hand, the 6284$\angstrom$ DIB gives a better value of the correlation coefficient but the behaviour of the points in the plot suggests that we should be careful about this result (Fig.~\ref{figtest3}, bottom right panel).

There seem to be broad correlations for the 4430$\angstrom$ and 5797$\angstrom$ DIBs in the 5780$\angstrom$ double-trend region. As the name of the region suggests, the 5780$\angstrom$ DIB shows two different trends (Fig.~\ref{figtest4}, upper right panel). Although uncertainties may play a certain role in this result, it is very unlikely that a random error would produce such clear differences between two parts of the plot. Moreover, we note that the width of this region is relatively small (compared to the other regions). Comparing with the latitude plot, it seems that the most likely explanation is that this trend is the result of the zero-point shift. It is possible, that this could be corrected in a similar way as we corrected the data for 5797$\angstrom$ band, but as was mentioned before, in this case the value of the shifted zero-point is not easy to find.

\begin{figure}
  \resizebox{\hsize}{!}{\includegraphics{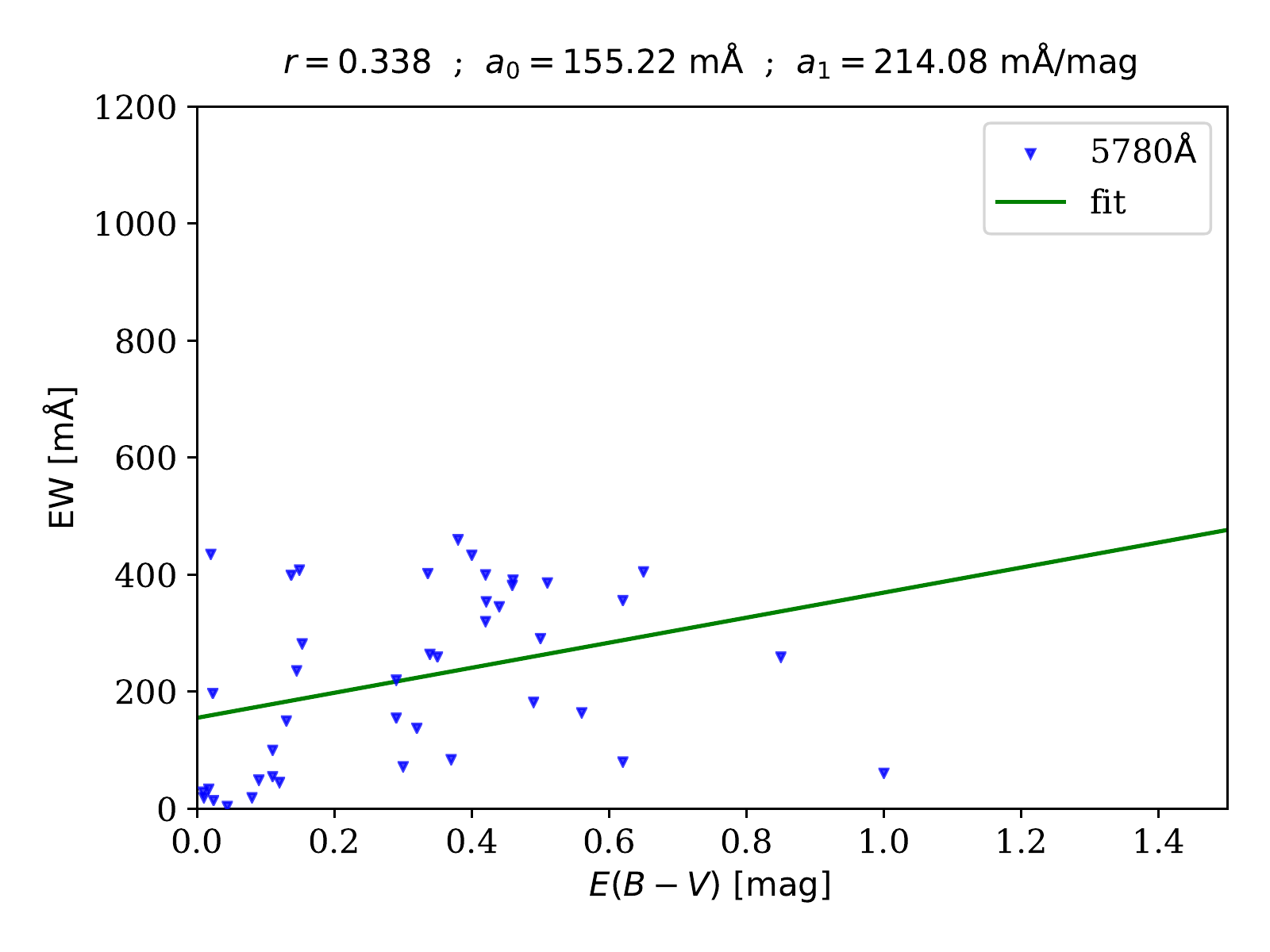}}
  \caption{The correlation between the strength of the 5780$\angstrom$ DIB and the colour excess $E(B-V)$ in the scatter-field region.}
  \label{fig18}
\end{figure}

The scatter-field region shows essentially the same picture for each of the DIBs. The correlation is either missing or the separation by longitudes needs to be finer in this region (Fig.~\ref{fig18}, Fig.~\ref{figtest5}). This behaviour is quite unique in the picture of the DIBs.

In the region of the unknown trend, we see a significant improvement of the correlation coefficients for the 4430$\angstrom$ and 6284$\angstrom$ DIBs (Fig.~\ref{figtest6}). For the other two DIBs, the correlation is much weaker, although still apparent in the plots. The name of the region was chosen based on the high value of the correlation coefficient of the 6284$\angstrom$ DIB (although slight improvement can be seen also in the case of the 4430$\angstrom$ DIB), which seems striking especially when we compare these results with those of the next region.

\begin{figure}
  \resizebox{\hsize}{!}{\includegraphics{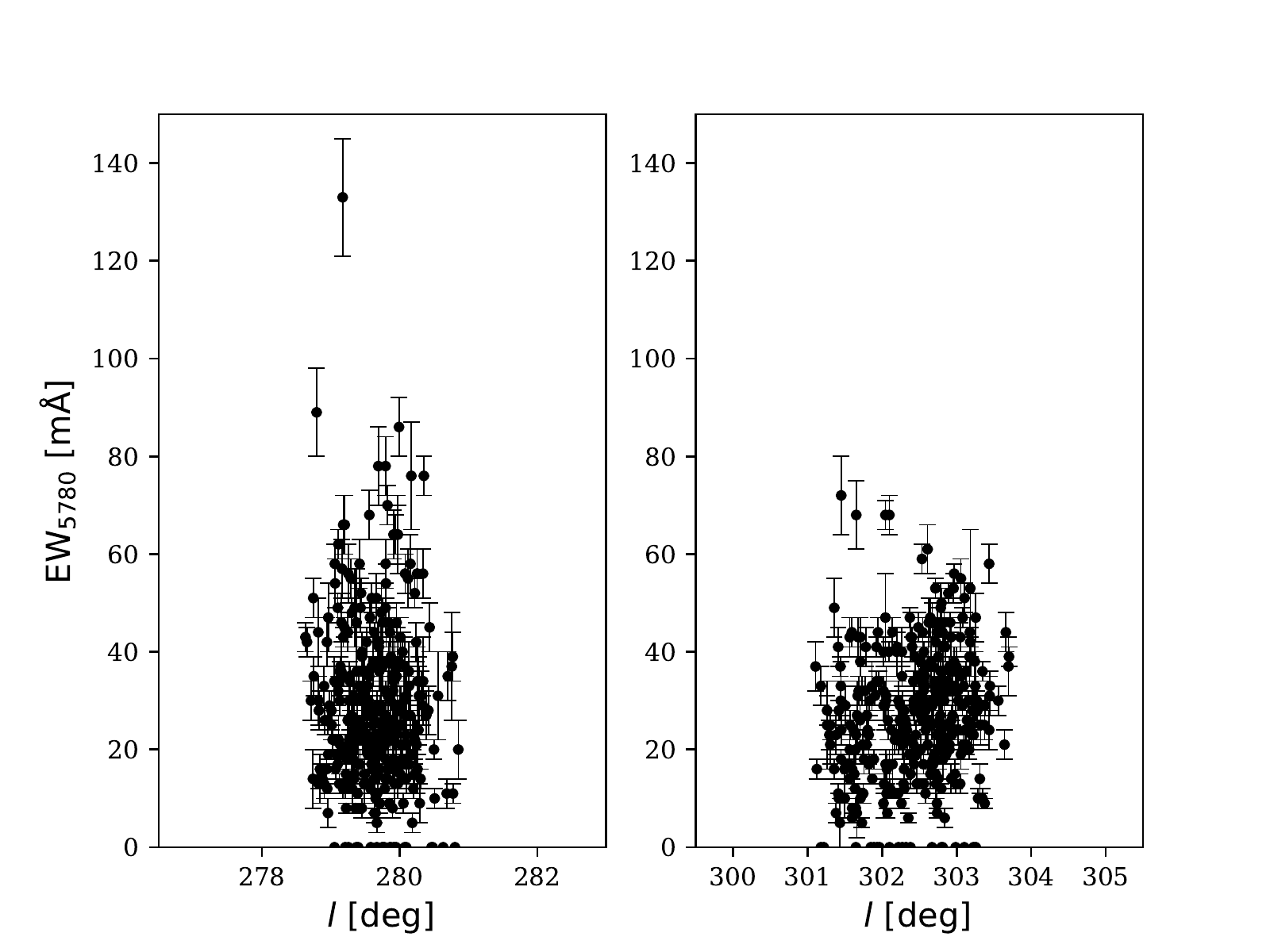}}
  \caption{Equivalent widths of the foreground 5780$\angstrom$ DIB in the lines of sight towards the LMC (left) and the SMC (right), taken from \citet{2015MNRAS.454.4013B}. The low value of the strength of the DIB should be noted. There is an interesting behaviour of the dependence in the direction towards the SMC, with an apparent rising towards the higher values of $l$.}
  \label{figBailey}
\end{figure}

Finally, the wide-minimum region shows that the correlation for 4430$\angstrom$ and 6284$\angstrom$ DIBs improves. For the 5797$\angstrom$ DIB, there are simply not enough points to say anything about its behaviour. On the other hand, we have enough data for the 5780$\angstrom$ DIB which shows a weaker correlation with the reddening (Fig.~\ref{figtest7}, upper right panel). However, when we look at the plot we see that the values of EWs are very low -- it is a local minimum. It is possible that the correlation for the 5780$\angstrom$ DIB cannot be found at such low values. Although we were able to find a trend in the first-minimum region, we note that the DIBs are much stronger there and the number of measurements is lower. To analyse this further, we checked with the data from \citet{2015MNRAS.454.4013B}, who provided measurements of the 5780$\angstrom$ and 5797$\angstrom$ bands towards the Large Magellanic Cloud (LMC) and Small Magellanic Cloud (SMC). They were able to resolve DIBs originating from both, the foreground ISM of our Galaxy and the Magellanic Clouds themselves. The data fit well together with the longitudinal plots of the 5780$\angstrom$ band in the wide-minimum region. Moreover, when we look more closely (Fig.~\ref{figBailey}), we can clearly see some sort of trend going on at larger scale, which is most likely connected to the interstellar medium of our Galaxy. However, we need to point out the important fact that Magellanic Clouds are located at higher Galactic latitudes than most of the other target stars used in this work. Finally, it is possible to compare our results from the wide-minimum region with the data from \citet{2016A&A...585A..12B} and \citet{2019NatAs...3..922F} which show basically the same result.

\section{Comparison With Extinction Maps}

\begin{figure}
  \resizebox{\hsize}{!}{\includegraphics{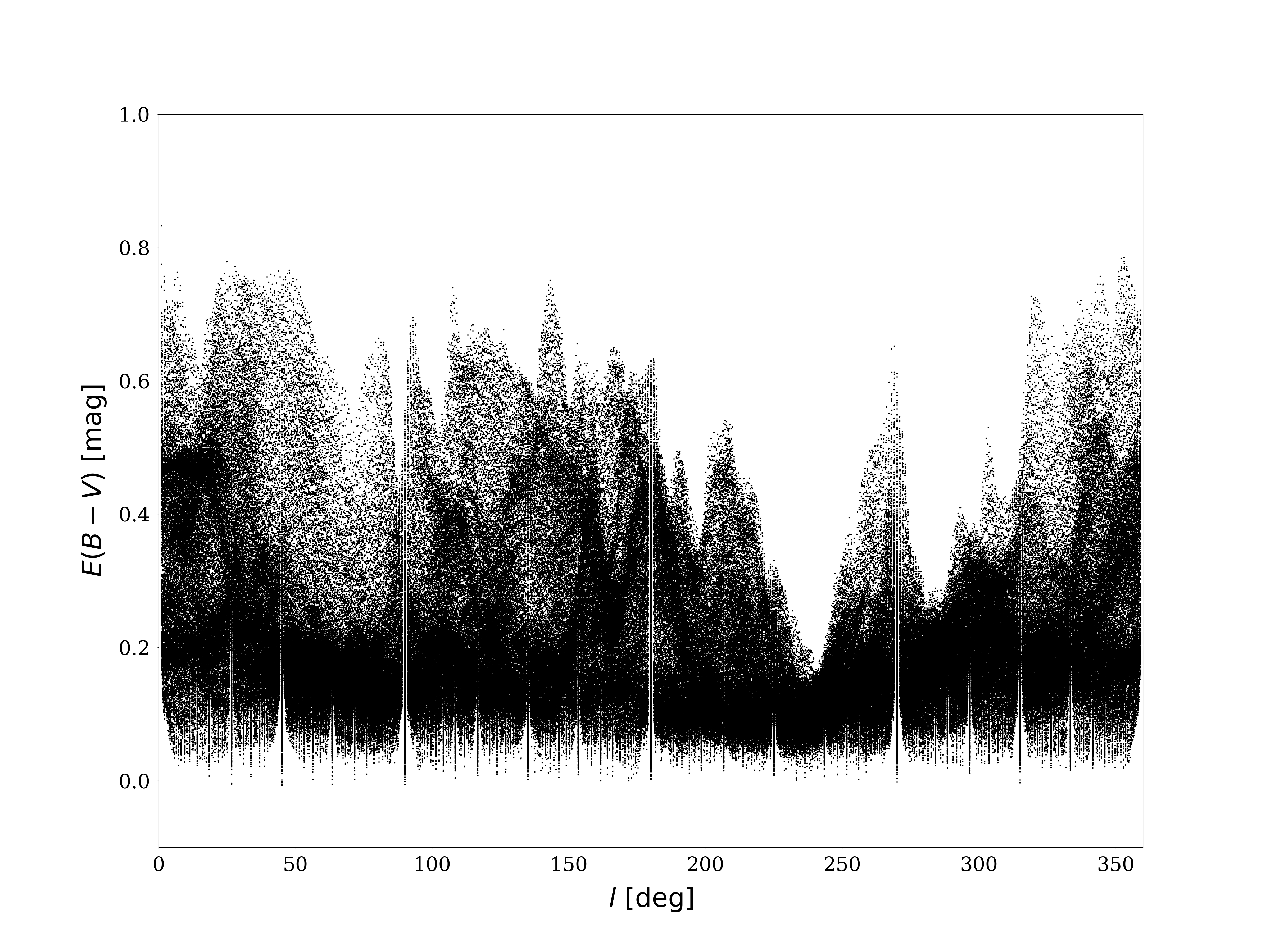}}
  \caption{Colour excess $E(B-V)$ in different lines of sight, taken from \citet{2018yCat.2354....0G}. There appears to be slight resemblance of the upper outline of the distribution of points with the one that we see in Fig.~\ref{fig03}.}
  \label{figG1}
\end{figure}

It might be helpful to compare the result we derived with the available extinction maps. We have used the data from VizieR -- one of the best maps, that are available there, are from \citet{2018yCat.2354....0G}. Since these maps are already restricted in the Galactic height ($-600$ pc $< Z < 600$ pc), we only require the $A_V$ or $E(B-V)$, $X$ and $Y$ data. From these, we can derive the Galactic longitude plot of the extinction -- see Fig.~\ref{figG1}. With careful inspection of the comparison with the longitudinal plots of the four bands we see that the DIBs seem to follow the pattern of the interstellar extinction, at least on the global scale. However, there are a few parts of the extinction map at which we should take a closer look.

\begin{figure}
  \resizebox{\hsize}{!}{\includegraphics{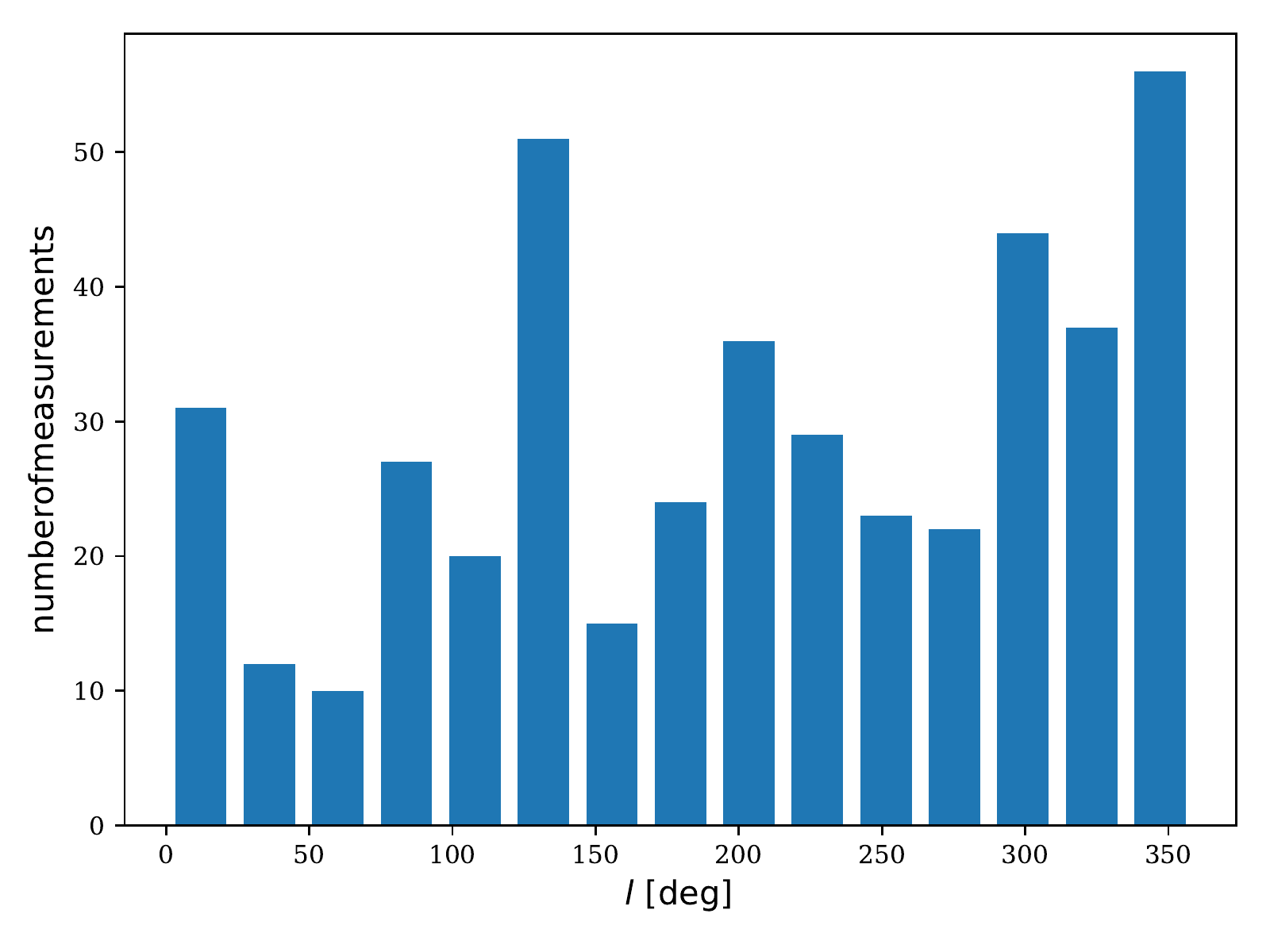}}
  \caption{Histogram which shows us the number of non-zero measurements of the equivalent widths for the 5780$\angstrom$ DIB.}
  \label{figHH}
\end{figure}

\begin{figure}
  \resizebox{\hsize}{!}{\includegraphics{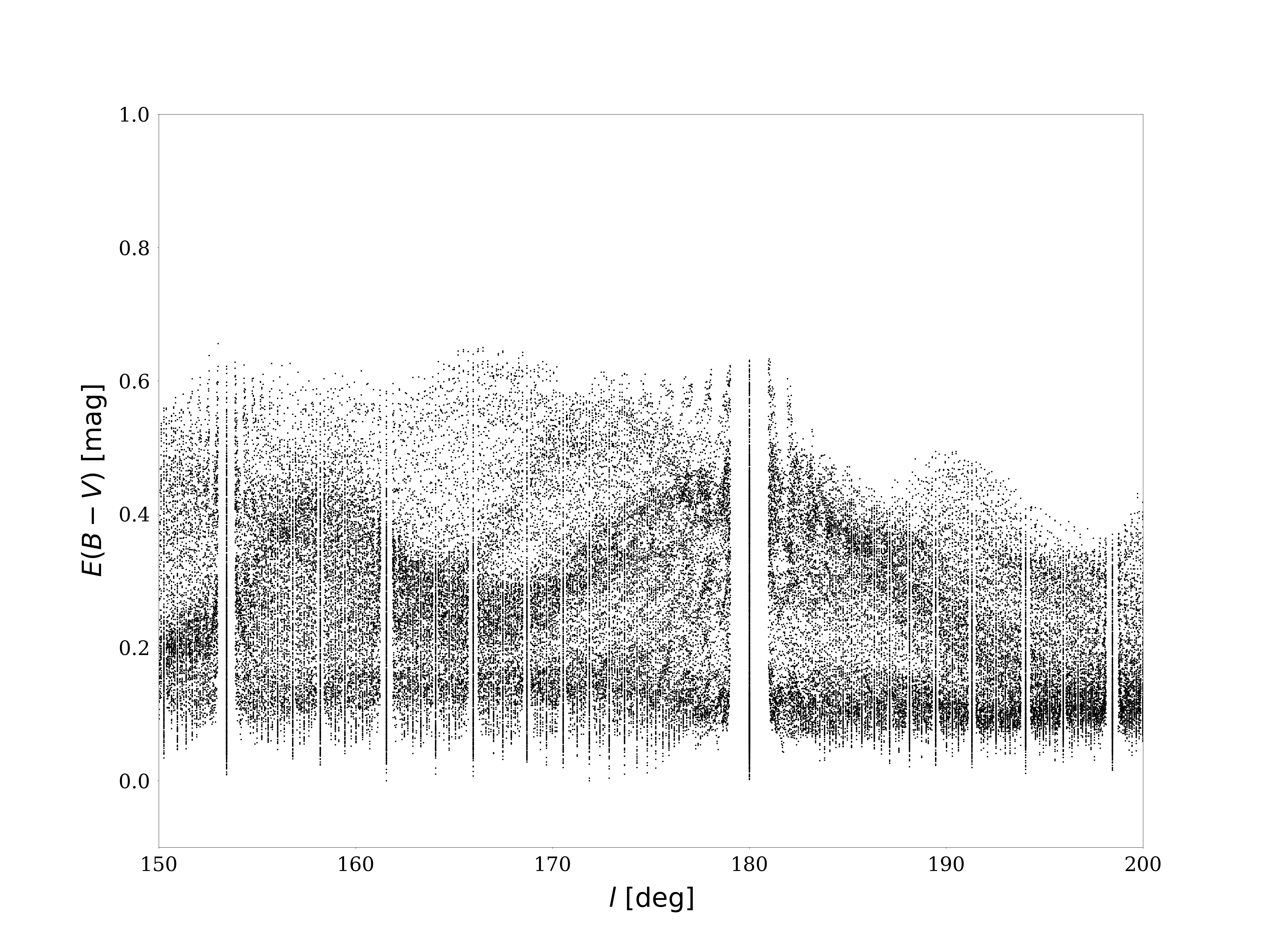}}
  \caption{Appearance of different distributions of reddening in the lines of sight which coincide with the scatter-field region. Presence of such structures is explained by the existence of dense interstellar medium which is related to the nearby molecular clouds between us and the target stars.}
  \label{figG2}
\end{figure}

\begin{figure}
  \resizebox{\hsize}{!}{\includegraphics{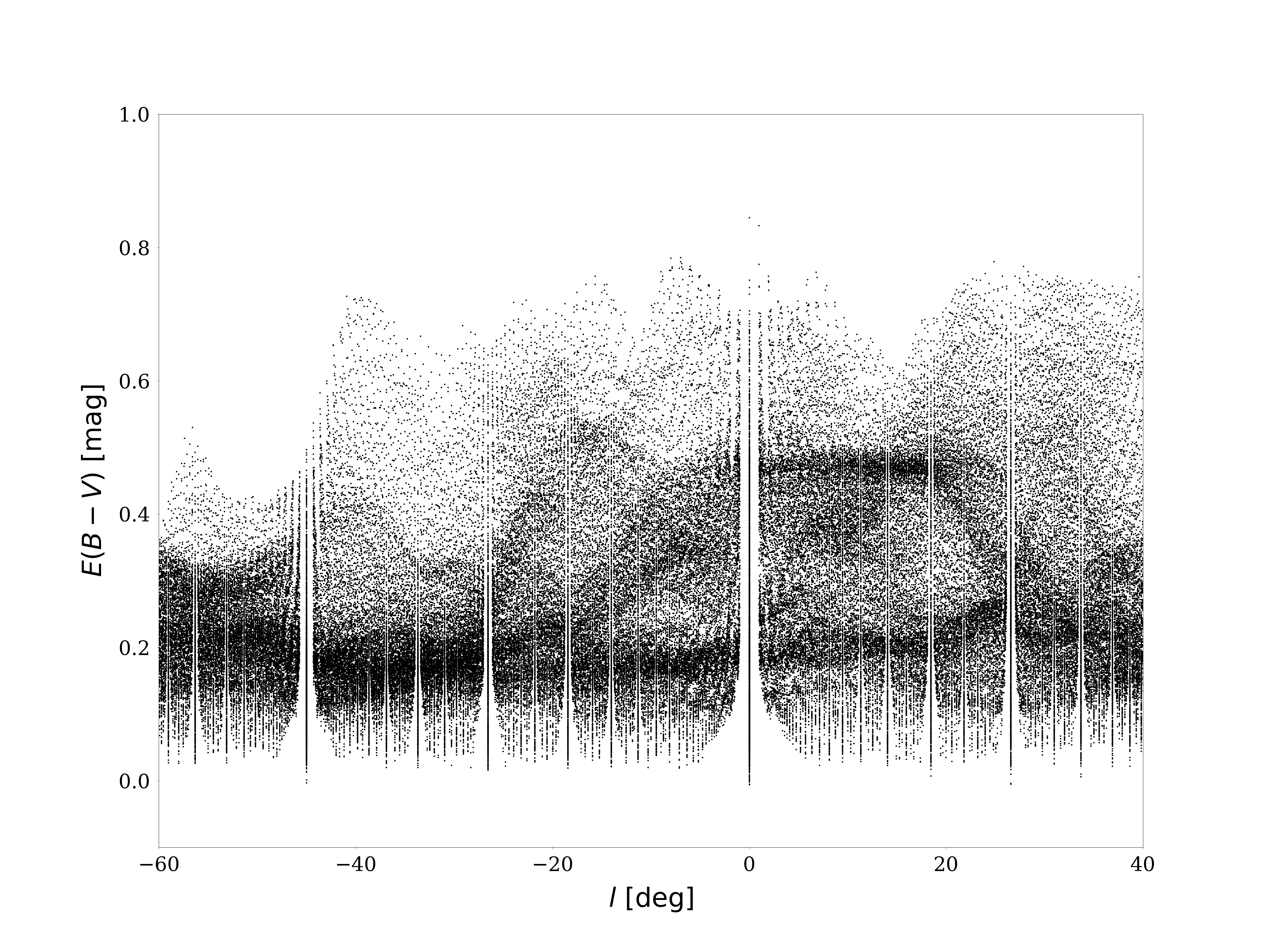}}
  \caption{Appearance of different distributions of reddening in the lines of sight which coincide with the Galactic-central region. Correlations between the DIBs and the colour excess is quite good, despite the presence of regions of denser interstellar medium in these directions (towards the Galactic centre).}
  \label{figG3}
\end{figure}

First of all, the region between $l=40^\circ$ and $l=75^\circ$ does not quite match with the DIBs. This may be due to the fact that we do not have many observations in these lines of sight -- this can be seen in the histogram displayed in Fig.~\ref{figHH}. It must be noted that the behaviour in the DIB maps and in the extinction corresponds very well in the Galactic-central, first-peak and 5780$\angstrom$ double-trend regions.

The wide-minimum region seems to match the behaviour in the extinction map quite well. On the other hand, when we compare the relative strength of DIBs in the wide-minimum and the first-peak regions with the ratio in extinction, the strength of the DIBs appears to fall down a bit more quickly in this region than the extinction map would suggest, although the EW uncertainties are too high and number of measurements too low for us to be sure.

Finally, we would like to discuss the scatter-field region. We have seen that for the DIBs it is probably impossible to find a correlation with the colour excess here. If we take a closer look at this region in the extinction map (Fig.~\ref{figG2}), we can see that there is a very complicated and strong structure on top of a bottom structure which is the continuation of the behaviour at smaller and larger longitudes. It is very likely that this has some relation to the number of nearby molecular clouds located in this region, like the prominent California \citep{2009ApJ...703...52L}, Orion \citep{1994A&A...281..517J}, Perseus \citep{2018ApJ...865...73O} and Taurus \citep{1991MNRAS.252..234A} clouds. It was shown that DIBs are seen weaker towards dark regions \citep[e.g.][]{1974ApJ...194..313S} and stronger towards regions with strong UV radiation fields \citep[e.g. Orion nebula,][]{1994A&A...281..517J}. The only DIB which appears to be an exception (in our set of bands) is the 6284$\angstrom$ DIB which displays a very similar strength towards both, the Orion nebula and Taurus dark clouds \citep{1994A&A...281..517J}. The results of our analysis show that all of the four DIBs seem to be less correlated with the colour excess in the "scatter-field" region when compared with the original (unseparated) data set. We would like to point out that this does not contradict the results of the previous works. Let us assume that the observed lines of sight are distributed uniformally across the sky and that the absolute dimension for different clouds are almost the same. It is clear that we should get much more data points related to the clouds which are closer -- this is due to the fact that the apparent size of a cloud depends on the distance from this cloud. Therefore, it is less likely that more than one line of sight passes through a distant cloud than it is in the case of clouds which are closer. Let us now analyse the case of getting multiple measurements related to a single cloud. We will assume that for two distinct segments of a cloud we can find identical values of the correlation coefficient. The conditions (such as temperature, UV radiation field, density) in the Orion nebula, and other similar clouds, change as we look toward different regions of the whole complex -- generally, the relation between the EWs and the $E(B-V)$ should vary (in terms of different parameters of the fitting function) as we study different parts of the cloud. The net effect for the whole cloud would then be a much lower value of the overall correlation coefficient when compared with the individual segments.

A second effect may also take place with varying contributions to the scatter in the plots. It is a result of the fact that, given a line of sight, the number of absorbing particles depends on the distance where we look. When looking further away and comparing with shorter distances, it can be expected that the column densities are going to be dominated mostly by the diffuse parts of the ISM, if such regions are present in the line of sight.

There is only one other region which displays a similar structure as can be seen in Fig.~\ref{figG2} -- the one at the centre of the Galaxy (see Fig.~\ref{figG3}). It is puzzling why we see an improvement in the correlation coefficients of the 4430$\angstrom$ and 5780$\angstrom$ DIBs when we would expect the opposite to happen (as for the other two DIBs), similarly to the scatter-field region. It is possible that this can be the result of different distances of the observed stars. However, when looking at Fig.~\ref{figH1} and Fig.~\ref{figH2}, we see that these two regions are quite similar from the statistical point of view.

\section{Discussion}

Overall, we have found the following new results:
\begin{itemize}
\item There appears to be a structure in the Galactic coordinate $x$ seen in the 5780$\angstrom$ (and maybe 6284$\angstrom$) DIB within $|x|<250$ pc. The structure appears to be non-symmetric and is not of statistical origin, based on the relatively small errors of EWs and a good sampling of data points.
\item The correlation coefficients between the EWs of the individual DIBs and the colour excess $E(B-V)$ should change if we split the total data into longitudal regions. This was done with the help of the overall longitudal plot and resulted in different changes for different DIBs (most noticable is the comparison of the 5780$\angstrom$ and the 6284$\angstrom$ DIBs). The region defined by the interval of longitudes $150^\circ < l < 200^\circ$ displays a significantly lower value of correlation coefficients for all of the studied DIBs.
\item Scatter seen in some of the regions is most likely caused by the presence of nearby molecular clouds. This is a result of the fact that physical conditions change as we look at different parts of a cloud. Another effect which must be considered is coming from different contribution of the diffuse regions along the line of sight when the line of sight crosses a cloud.
\item For the 5780$\angstrom$, 5797$\angstrom$ and 6284$\angstrom$ DIBs, the region located in the interval $250^\circ < l < 300^\circ$ appears to contain a smaller amount of carriers (inferred from the lower typical values of EWs) when compared with the other regions. The result for the 4430$\angstrom$ DIB is most likely biased in the Galactic plane by the lack of observations in the fourth quadrant.
\end{itemize}

We have shown that with the use of publicly available data, we can construct maps that can be readily used to extract information about DIBs. We have seen that the strength of the DIBs is centred around the Galactic plane while, on the other hand, there is a much more complicated structure in the longitudinal plots. Moreover, it appears that the 4430$\angstrom$ DIB displays a different behaviour in these plots than the other three studied DIBs.

With the use of the recently released data from Gaia-DR2, we were able to get the distances towards our target stars from \citet{Bailer2018} (if available) with, for the most part, relatively low uncertainties. This made it possible to construct plots which show different behaviour of the EWs of the DIBs in the individual rectangular Galactic coordinates. It appears that there is a small-scale structure in the near vicinity of the Sun in the interval of $x$-values ($-250$ pc, $250$ pc). Due to the lack of data in the negative $y$ direction, we cannot be certain whether there is a structure in the plots of this coordinate. Finally, we have also shown very simple maps of the different DIBs in the Galactic $x$-$y$ plane. These show that we can see that the strength of a DIB may significantly vary towards different lines of sight even in the case when the observed stars have similar distances. Making maps such as those using much larger amount of data points could point towards the global distribution of the carriers around the Sun (within 1 kpc).

Using the results from the longitudinal plots and the fact that the correlation plots between the EWs and the reddening show very wide spread, we "detrended" the plots to several subregions which all display different behaviours. Most interesting are the "Galactic-central" and "scatter-field" regions. We see multiple trends in these two regions even when looking at the extinction maps. We argue that this behaviour is explained by the complicated structure of the ISM towards the Galactic centre and the presence of a nearby complex of giant molecular clouds in the "scatter-field" region. However, the results shown in the "detrended" longitudinal plots are somewhat inconsistent -- while the presence of molecular clouds can likely be the source of the scatter that we see in the related plots, there are obvious trends in the "Galactic-central" region which seems to point to the fact the complicated nature of the ISM in these lines of sight (despite containing molecular clouds) does not affect the behaviour of the DIBs as much as the presence of a big complex of molecular clouds.

However, we can expect that there are clouds present also in other lines of sight than towards the "scatter-field" region, and these do not seem to be a source of a significant scatter in other plots. We suspect that the distances toward the probed clouds, varying conditions throught the clouds, and the distribution of lines of sight on the sky are the reason behind this different behaviour. Moreover, distance together with the column density of the carriers can have another effect -- looking in the same direction (same column density) gives a different number of carriers (and therefore different EWs) at various distances. Therefore, looking in a direction of a cloud which is, for example, 300 pc away will result in carriers which are close to (or within) the cloud to be a more dominant source of absorption than it would be in the case of this cloud being more distant. Due to the lack of knowledge about the distribuition and the structure of the carriers, it is impossible to estimate how important this effect can become.

To search for other clues regarding the "Galactic-central" region, we looked in the literature for works focused on the DIBs and molecular or dark clouds. According to \citet{1991MNRAS.252..234A}, there could be a strong connection between some of the DIBs as they vary across the cloud and their strength in general seems to be very low in the sight-lines towards dark clouds. Moreover, \citet{1994A&A...281..517J} were unable to find a good correlation between the equivalent widths of DIBs and the extinction in the Orion molecular cloud. Our results appear to be consistent with their findings. Finally, studies of the Sco OB2 near the Galactic centre \citep{2011A&A...533A.129V} show that the correlation of the DIBs with the reddening is, on average, very similar to the "general" correlation in the diffuse interstellar medium. This is slightly in conflict with our findings since we have shown that there is not going to be a "general" correlation, but the idea of averaging of the correlations towards the regions near the Galactic centre could explain why we do not see scatter in the EW vs $E(B-V)$ plots. This can be supported by the fact that the total correlation differs only slightly from the "detrended" correlation in the "Galactic-central" region. However, we need to point out that this region in our study is still very broad compared to the one in the last mentioned work.

We also point our attention to the "first-peak" region. We can clearly see a small improvement in the correlation between the equivalent width and the reddening as a result of "detrending". However, this is not true for all of the DIBs. This can either be explained by the uncertainties of the determination of the equivalent widths, or by the existence of an absolutely different behaviour of the DIBs in these lines of sight. Moreover, the strength of all of the four DIBs seems to reach maximum values in these directions. We cannot explain why this is so -- for comparison, we looked at the maps of the local interstellar medium from \citet{2014A&A...561A..91L} and \citet{2018A&A...616A.132L}. There are both, high-absorption (i.e. high-density) regions and diffuse medium regions in the interval of longitudes ($75^\circ$, $125^\circ$). This seems to point against our assumption that clouds are a source of a scatter in the plots of EWs against the colour excess but the clouds in this region are typically more distant. On the other hand, we also see in the maps from the mentioned works that there are several high-absorption regions (typically less distant than those seen in the "first-peak" region and located within $125^\circ < l < 150^\circ$) where we find that the correlation coefficients are somewhat lower for the 5780$\angstrom$ and the 5797$\angstrom$ DIBs. Again, this seems to indicate that the distances toward clouds (within the line of sight) seem to play some role in the study of the correlations.

The difference between the behaviour in the "double-trend" region and the "scatter-field" region can be explained by assuming the second effect of the distance mentioned above -- according to the map in \citet{2018A&A...616A.132L}, the "double-trend" region contains more high-absorption regions within 1 kpc than diffuse regions when compared with the "scatter-field" region.

In general, the "wide-minimum" region seems to contain the least amount of the high-absorption material. This may be the result of a lack of carriers in these regions of the ISM or an effect produced by an observational bias. Looking at the $x-y$ map of the 4430$\angstrom$, the small number of observations in the fourth quadrant (where large part of this "wide-minimum" region is located) will greatly affect the results. For the other three DIBs, the number of data points in the fourth quadrant are satisfactory. Although almost all observed stars are within a 500 pc radius, we can see larger values of EWs even at shorter distances. This points to the explanation that the carriers of 5780$\angstrom$, 5797$\angstrom$ and 6284$\angstrom$ DIBs are relatively underabundant in this region.

The complexity of the longitudinal plots was shown in greater detail in the data provided by \citet{Chen13} and \citet{2015MNRAS.454.4013B}. The OGLE stars in C13 tell us that there is an interesting behaviour towards the Galactic bulge which can be studied if we have enough data in $\Delta l=0.1^\circ$ intervals. On the other hand, some sort of behaviour on slightly larger scale appears to be also present in the lines of sight towards the SMC and LMC. With sufficient amount of high-quality data, it should be possible to connect small-scale structure across the whole sky into one distribution that could be used to analyse global properties of carriers of the DIBs.

Obviously, our "detrending" process reduced the plots to subplots which now consist of only small numbers of points. More data and observations are required to verify our findings. In the future, we aim to create an automatic procedure that would search the spectra available in the databases (for example, ESO database), search for the presence of the DIBs and determine their equivalent widths. This would give us more information to work with and we could improve the quality of the maps (especially with the help of Gaia-DR2). Furthermore, we could study the "wide-minimum", "Galactic-central" and "scatter-field" regions in much higher detail. It is quite possible that studies of the regions (on small and global scale), where the behaviour of the DIBs significantly differs from the "normal trends", will pinpoint the birthplace of the carriers of the DIBs.




\bibliographystyle{mnras}
\bibliography{bibref} 




\appendix

\section{Additional Plots}
Most of the plots used in the text are displayed in this section. Additionally, figures not mentioned in the text are presented as well. All of them display the relation between the equivalent widths of DIBs and Galactic coordinates or colour excess $E(B-V)$.

\begin{figure*}
  \resizebox{\hsize}{!}{\includegraphics{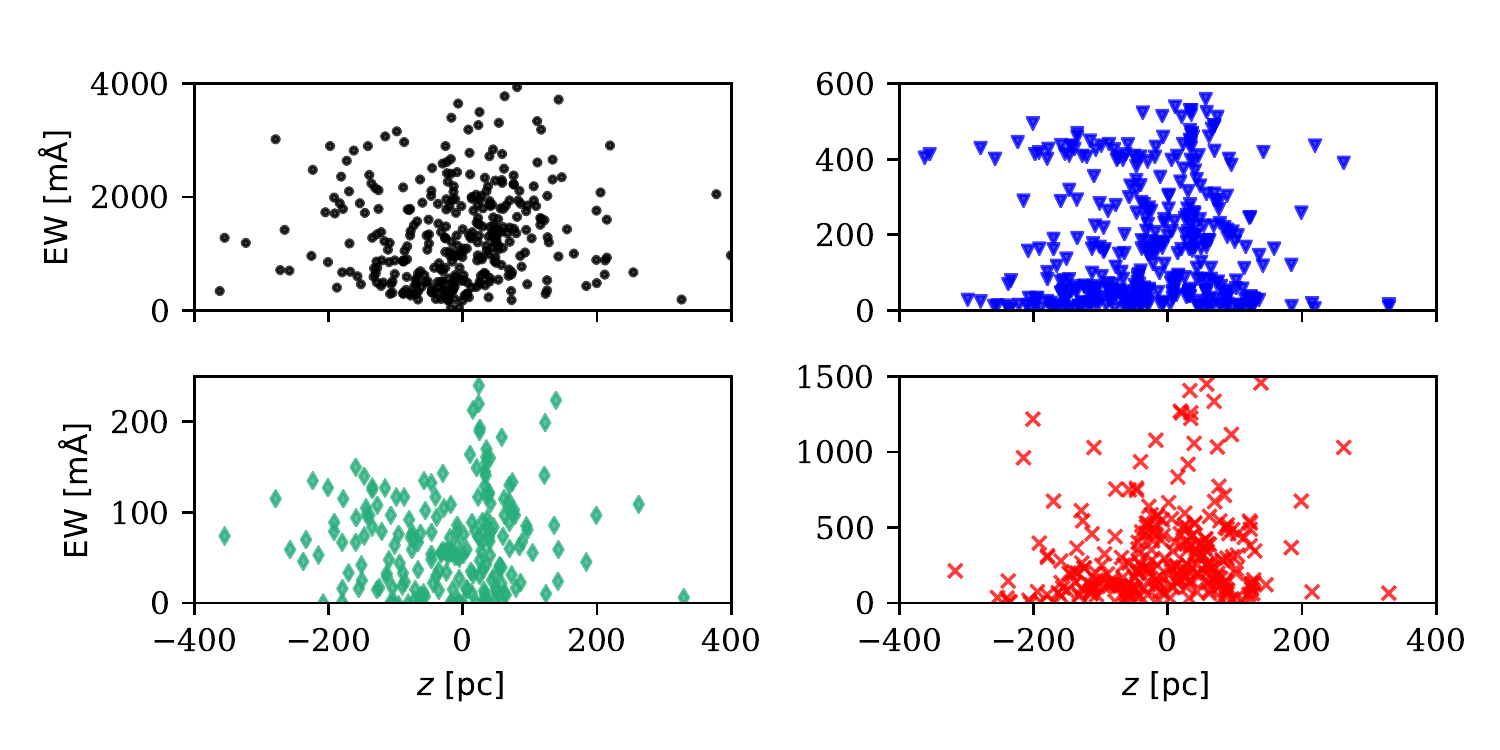}}
  \caption{Correlations between the strength of the DIBs and the Galactic $z$-coordinate. Size of the points depends on the strength of the DIB. Black dots show 4430$\angstrom$ DIB, blue triangles 5780$\angstrom$ DIB, teal diamonds 5797$\angstrom$ DIB and red crosses 6284$\angstrom$ DIB.}
  \label{figZ}
\end{figure*}

\begin{figure*}
  \resizebox{\hsize}{!}{\includegraphics{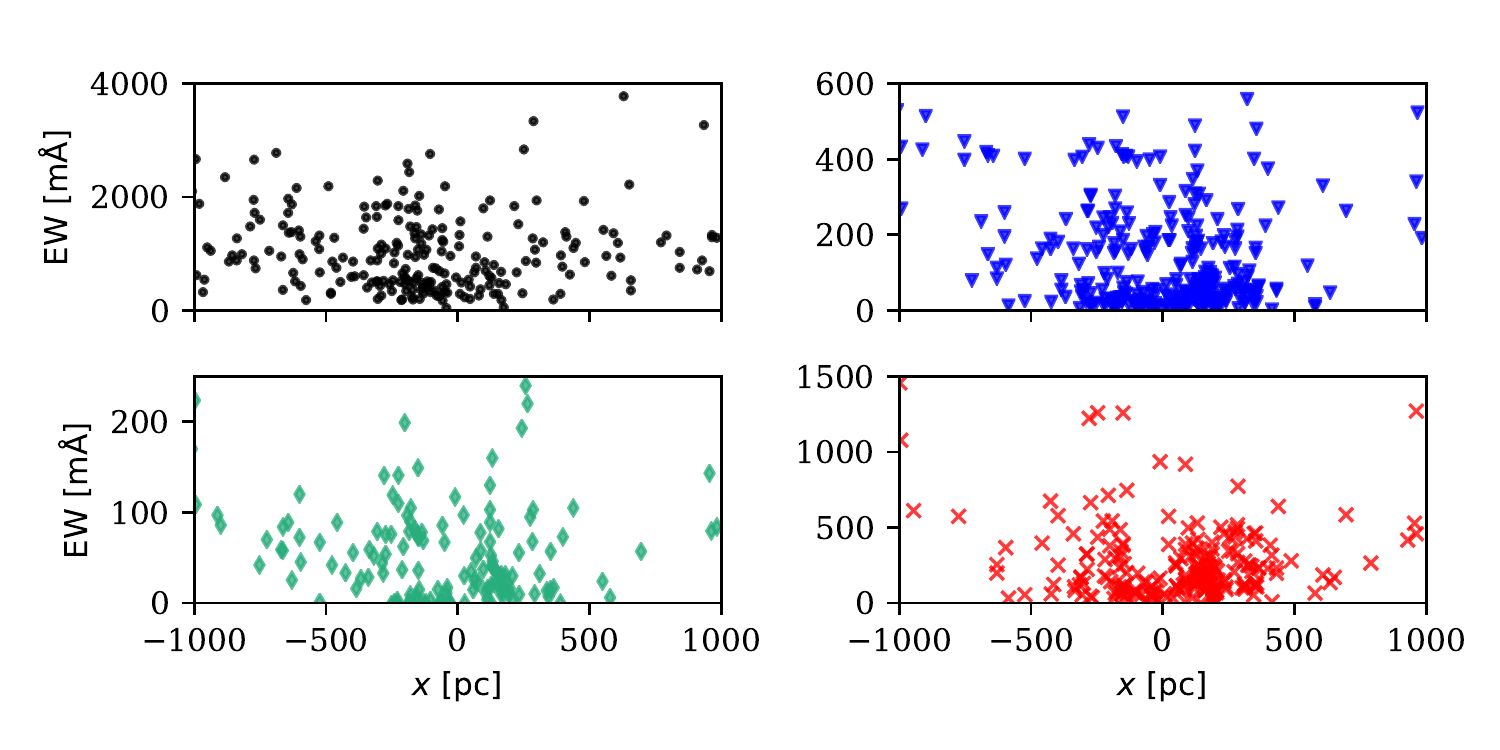}}
  \caption{Correlations between the strength of the DIBs and the Galactic $x$-coordinate. Size of the points depends on the strength of the DIB. Black dots show 4430$\angstrom$ DIB, blue triangles 5780$\angstrom$ DIB, teal diamonds 5797$\angstrom$ DIB and red crosses 6284$\angstrom$ DIB.}
  \label{figX}
\end{figure*}

\begin{figure*}
  \resizebox{\hsize}{!}{\includegraphics{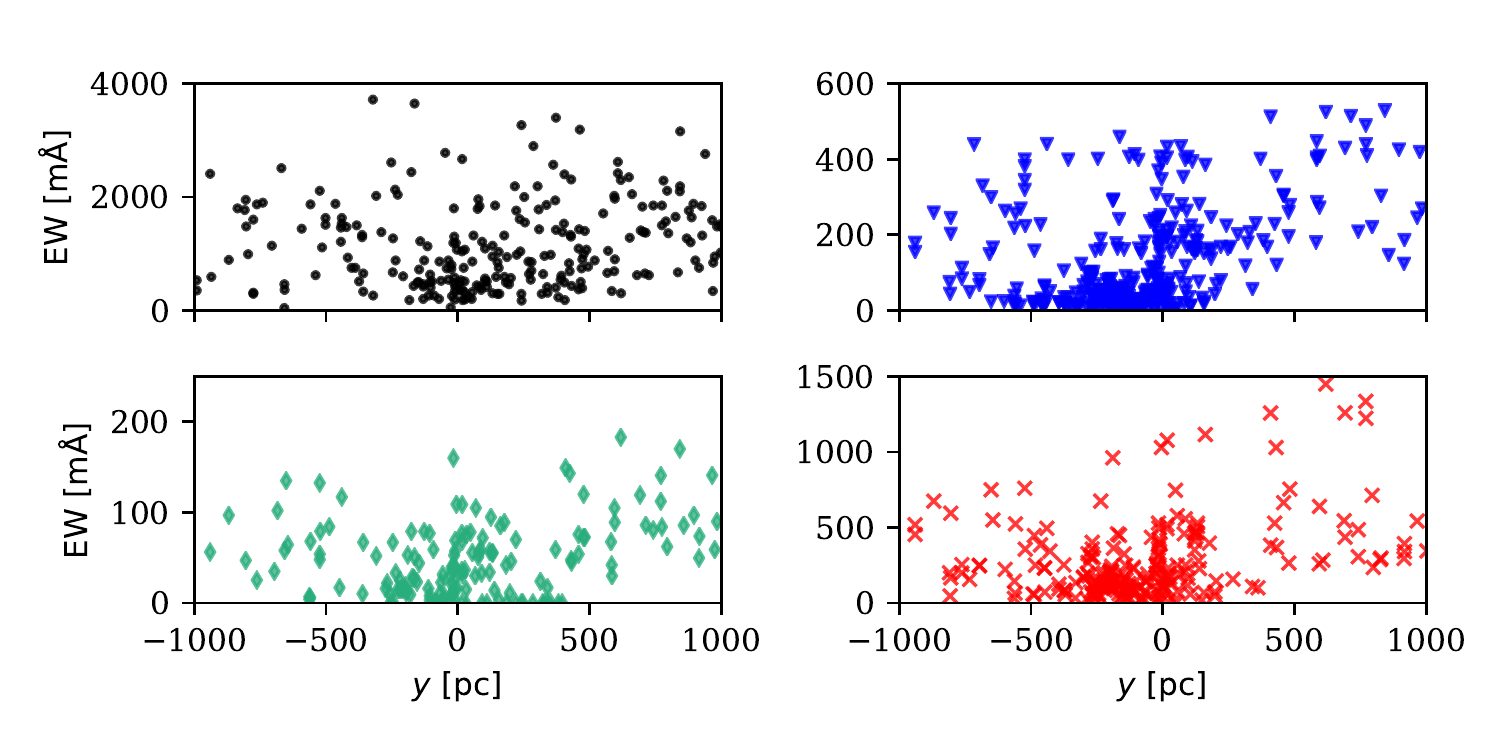}}
  \caption{Correlations between the strength of the DIBs and the Galactic $y$-coordinate. Size of the points depends on the strength of the DIB. Black dots show 4430$\angstrom$ DIB, blue triangles 5780$\angstrom$ DIB, teal diamonds 5797$\angstrom$ DIB and red crosses 6284$\angstrom$ DIB.}
  \label{figY}
\end{figure*}

\begin{figure*}
  \resizebox{\hsize}{!}{\includegraphics{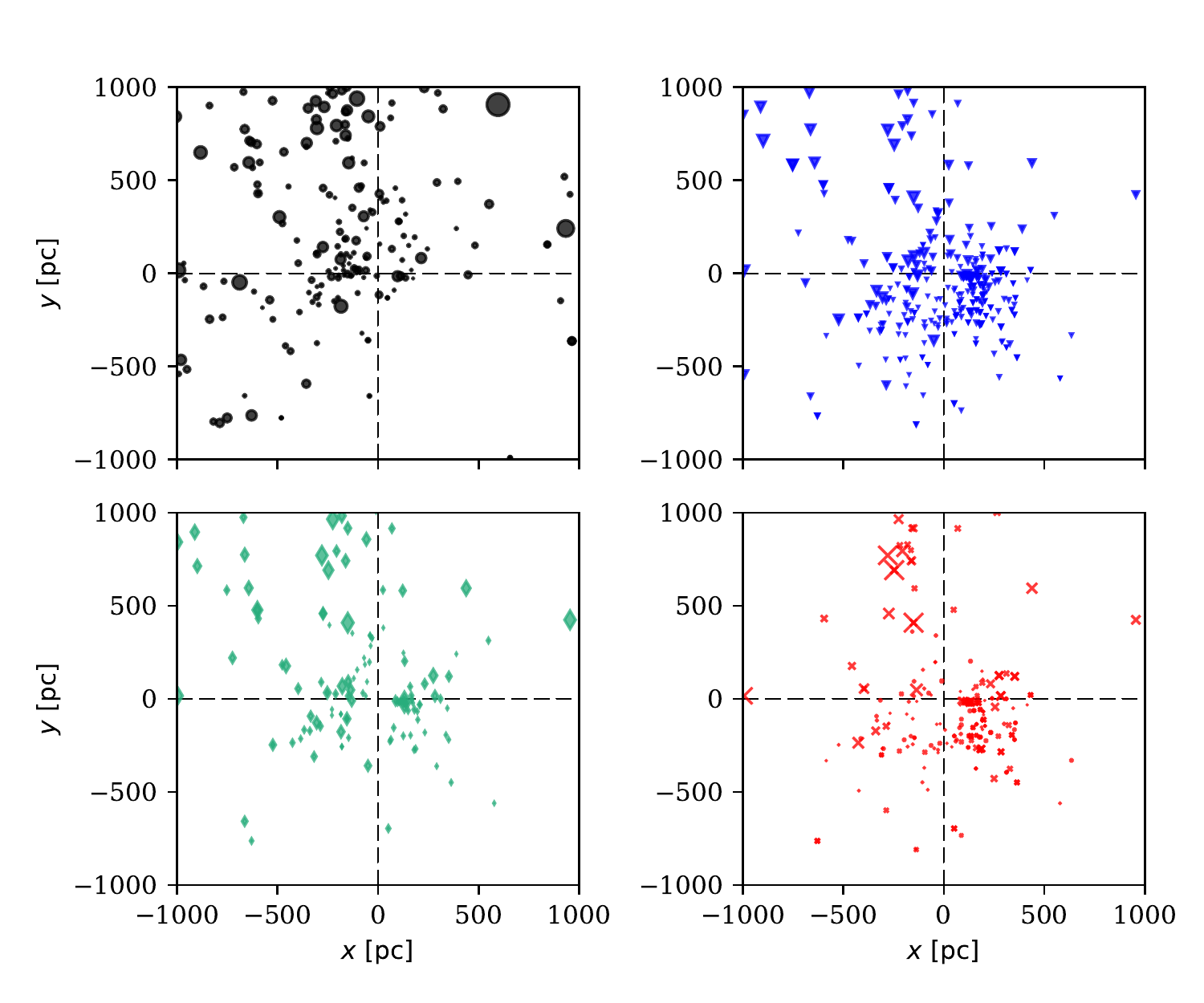}}
  \caption{Positions of the target star projected to the Galactic plane. Size of the points depends on the strength of the DIB. Black dots show 4430$\angstrom$ DIB, blue triangles 5780$\angstrom$ DIB, teal diamonds 5797$\angstrom$ DIB and red crosses 6284$\angstrom$ DIB.}
  \label{figXY}
\end{figure*}

\begin{figure*}
  \resizebox{\hsize}{!}{\includegraphics{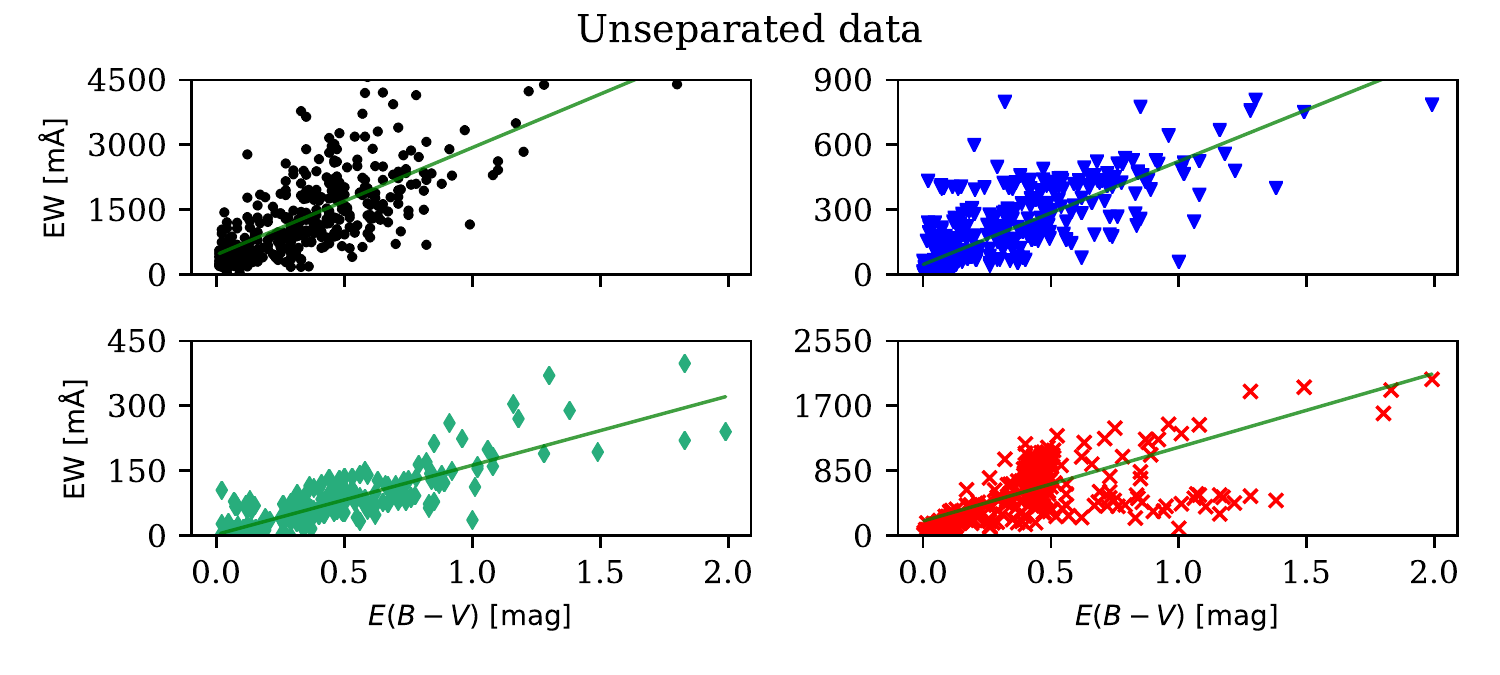}}
  \caption{Correlations between the strength of the DIBs and the colour excess $E(B-V)$ in the unseparated data. Green line represents a linear fit to the data with correlation coefficient $r$ and fit parameters $a_0$ and $a_1$. Black dots show 4430$\angstrom$ DIB ($r=0.675 \,,\, a_0=465 \,,\, a_1=2473$), blue triangles 5780$\angstrom$ DIB ($r=0.815 \,,\, a_0=47 \,,\, a_1=477$), teal diamonds 5797$\angstrom$ DIB ($r=0.850 \,,\, a_0=3 \,,\, a_1=160$) and red crosses 6284$\angstrom$ DIB ($r=0.704 \,,\, a_0=190 \,,\, a_1=968$).}
  \label{figtest0}
\end{figure*}

\begin{figure*}
  \resizebox{\hsize}{!}{\includegraphics{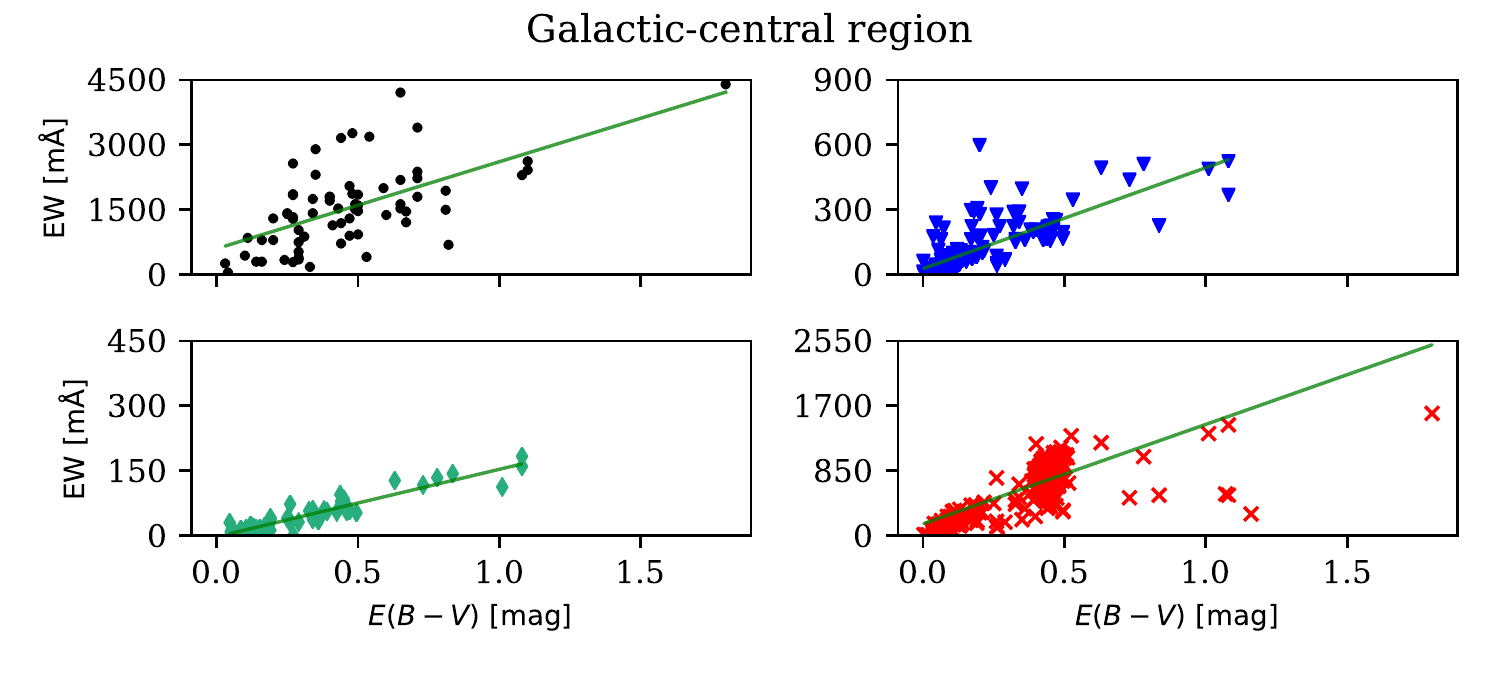}}
  \caption{Correlations between the strength of the DIBs and the colour excess $E(B-V)$ in the Galactic-central region. Green line represents a linear fit to the data with correlation coefficient $r$ and fit parameters $a_0$ and $a_1$. Black dots show 4430$\angstrom$ DIB ($r=0.612 \,,\, a_0=601 \,,\, a_1=2010$), blue triangles 5780$\angstrom$ DIB ($r=0.779 \,,\, a_0=28 \,,\, a_1=467$), teal diamonds 5797$\angstrom$ DIB ($r=0.942 \,,\, a_0=-3 \,,\, a_1=156$) and red crosses 6284$\angstrom$ DIB ($r=0.774 \,,\, a_0=154 \,,\, a_1=1302$).}
  \label{figtest1}
\end{figure*}

\begin{figure*}
  \resizebox{\hsize}{!}{\includegraphics{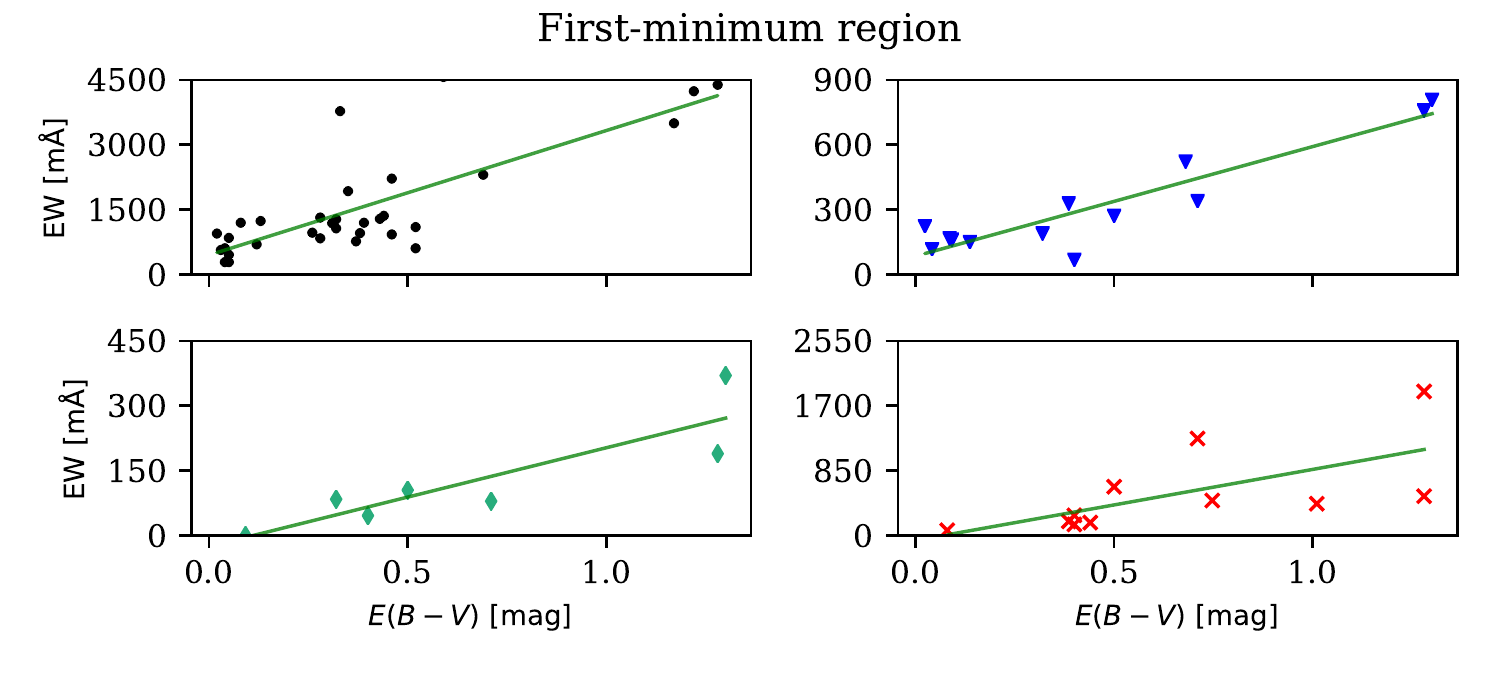}}
  \caption{Correlations between the strength of the DIBs and the colour excess $E(B-V)$ in the first-minimum region. Green line represents a linear fit to the data with correlation coefficient $r$ and fit parameters $a_0$ and $a_1$. Black dots show 4430$\angstrom$ DIB ($r=0.782 \,,\, a_0=453 \,,\, a_1=2879$), blue triangles 5780$\angstrom$ DIB ($r=0.923 \,,\, a_0=85 \,,\, a_1=507$), teal diamonds 5797$\angstrom$ DIB ($r=0.876 \,,\, a_0=-25 \,,\, a_1=228$) and red crosses 6284$\angstrom$ DIB ($r=0.653 \,,\, a_0=-64 \,,\, a_1=931$).}
  \label{figtest2}
\end{figure*}

\begin{figure*}
  \resizebox{\hsize}{!}{\includegraphics{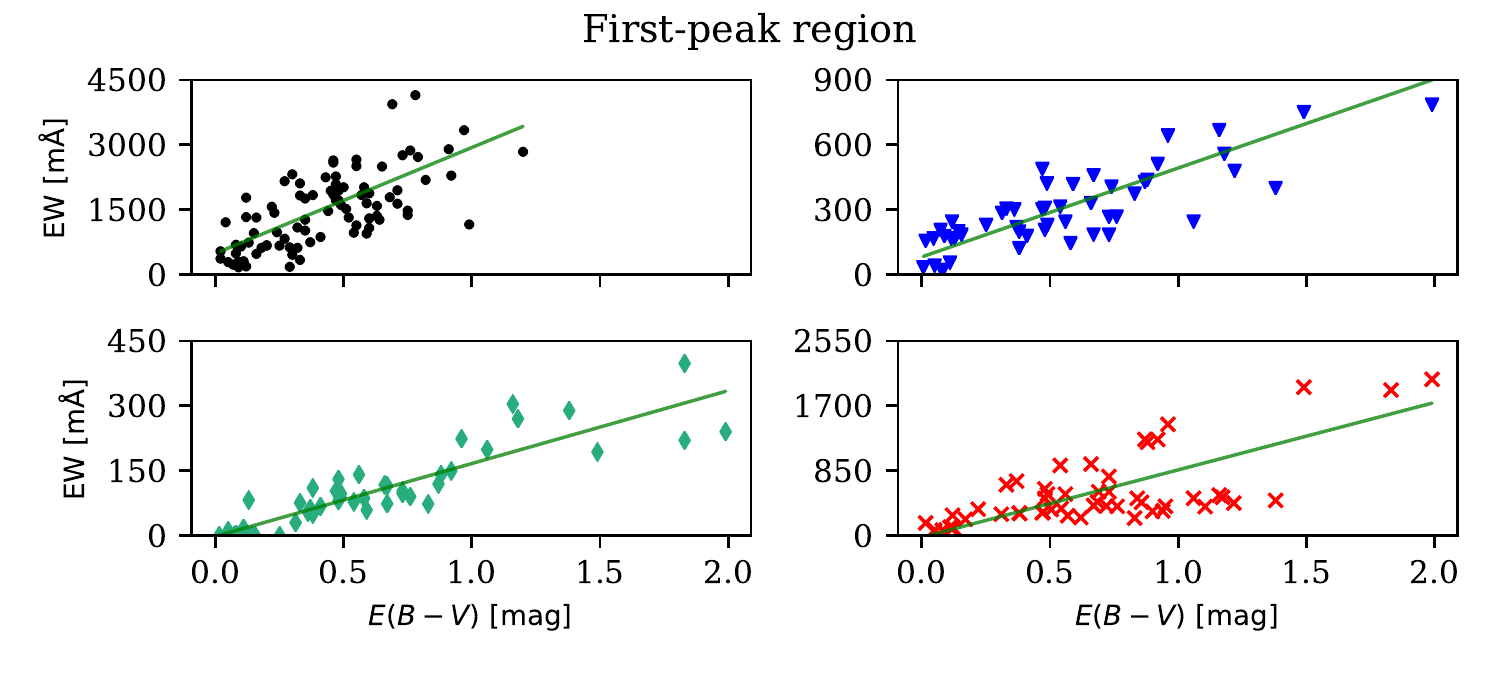}}
  \caption{Correlations between the strength of the DIBs and the colour excess $E(B-V)$ in the first-peak region. Green line represents a linear fit to the data with correlation coefficient $r$ and fit parameters $a_0$ and $a_1$. Black dots show 4430$\angstrom$ DIB ($r=0.633 \,,\, a_0=491 \,,\, a_1=2448$), blue triangles 5780$\angstrom$ DIB ($r=0.871 \,,\, a_0=82 \,,\, a_1=411$), teal diamonds 5797$\angstrom$ DIB ($r=0.897 \,,\, a_0=-2 \,,\, a_1=169$) and red crosses 6284$\angstrom$ DIB ($r=0.740 \,,\, a_0=-22 \,,\, a_1=883$).}
  \label{figtest3}
\end{figure*}

\begin{figure*}
  \resizebox{\hsize}{!}{\includegraphics{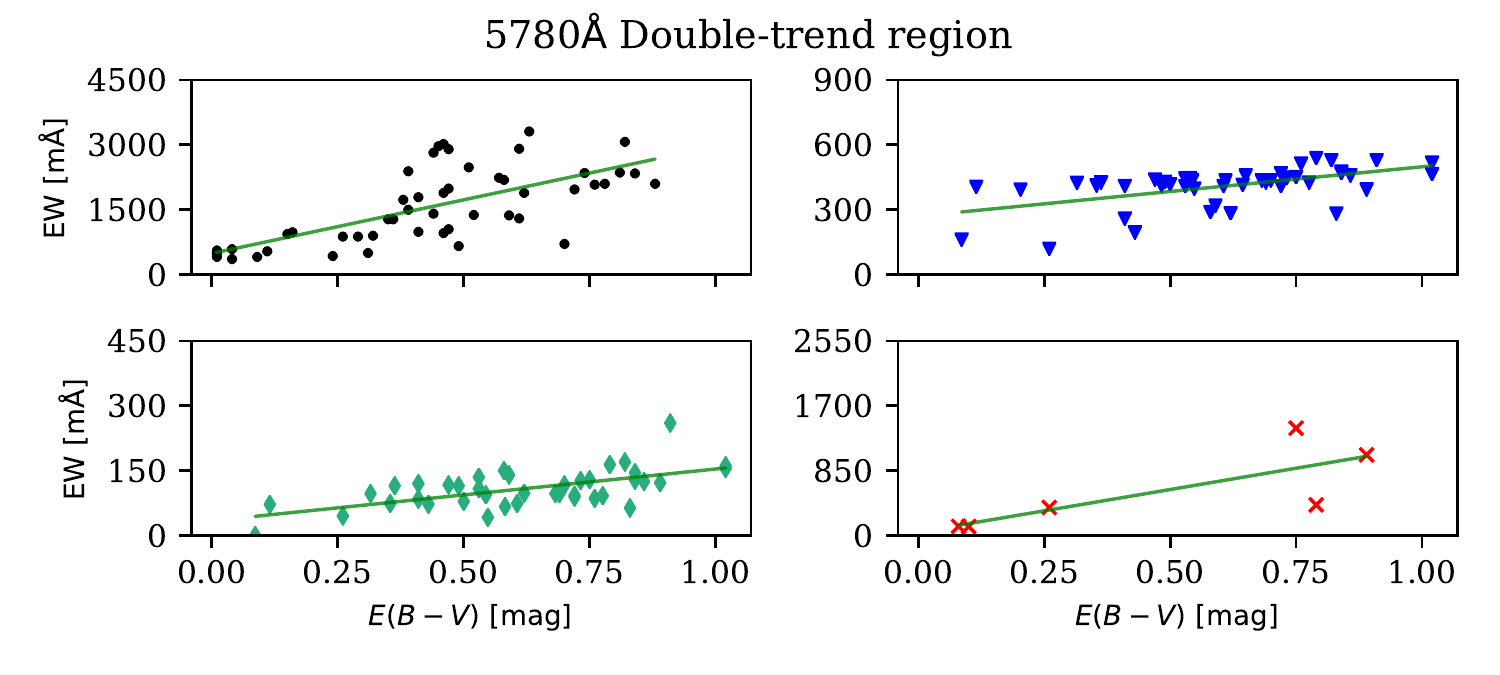}}
  \caption{Correlations between the strength of the DIBs and the colour excess $E(B-V)$ in the 5780$\angstrom$ double-tred region. Green line represents a linear fit to the data with correlation coefficient $r$ and fit parameters $a_0$ and $a_1$. Black dots show 4430$\angstrom$ DIB ($r=0.681 \,,\, a_0=492 \,,\, a_1=2476$), blue triangles 5780$\angstrom$ DIB ($r=0.553 \,,\, a_0=270 \,,\, a_1=229$), teal diamonds 5797$\angstrom$ DIB ($r=0.616 \,,\, a_0=34 \,,\, a_1=120$) and red crosses 6284$\angstrom$ DIB ($r=0.780 \,,\, a_0=45 \,,\, a_1=1116$).}
  \label{figtest4}
\end{figure*}

\begin{figure*}
  \resizebox{\hsize}{!}{\includegraphics{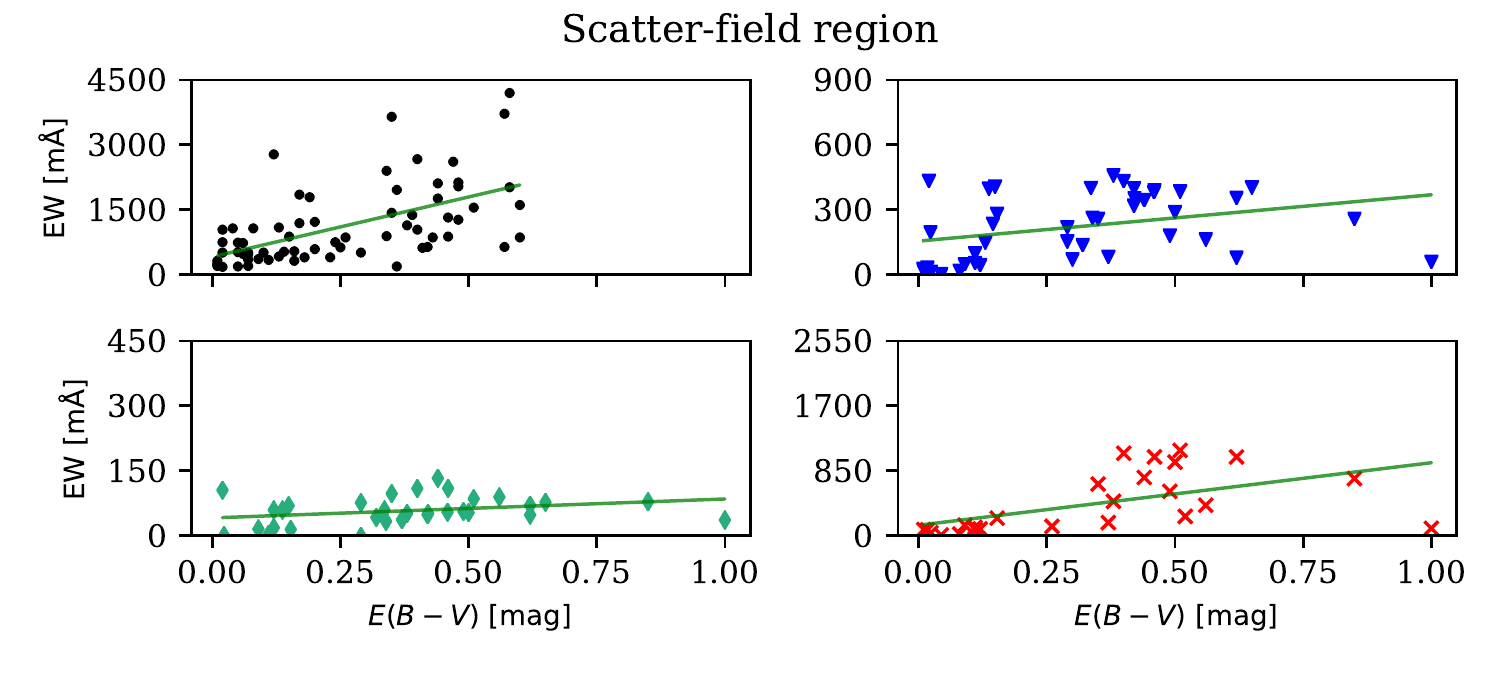}}
  \caption{Correlations between the strength of the DIBs and the colour excess $E(B-V)$ in the scatter-field region. Green line represents a linear fit to the data with correlation coefficient $r$ and fit parameters $a_0$ and $a_1$. Black dots show 4430$\angstrom$ DIB ($r=0.577 \,,\, a_0=412 \,,\, a_1=2769$), blue triangles 5780$\angstrom$ DIB ($r=0.338 \,,\, a_0=155 \,,\, a_1=214$), teal diamonds 5797$\angstrom$ DIB ($r=0.299 \,,\, a_0=41 \,,\, a_1=44$) and red crosses 6284$\angstrom$ DIB ($r=0.553 \,,\, a_0=135 \,,\, a_1=819$).}
  \label{figtest5}
\end{figure*}

\begin{figure*}
  \resizebox{\hsize}{!}{\includegraphics{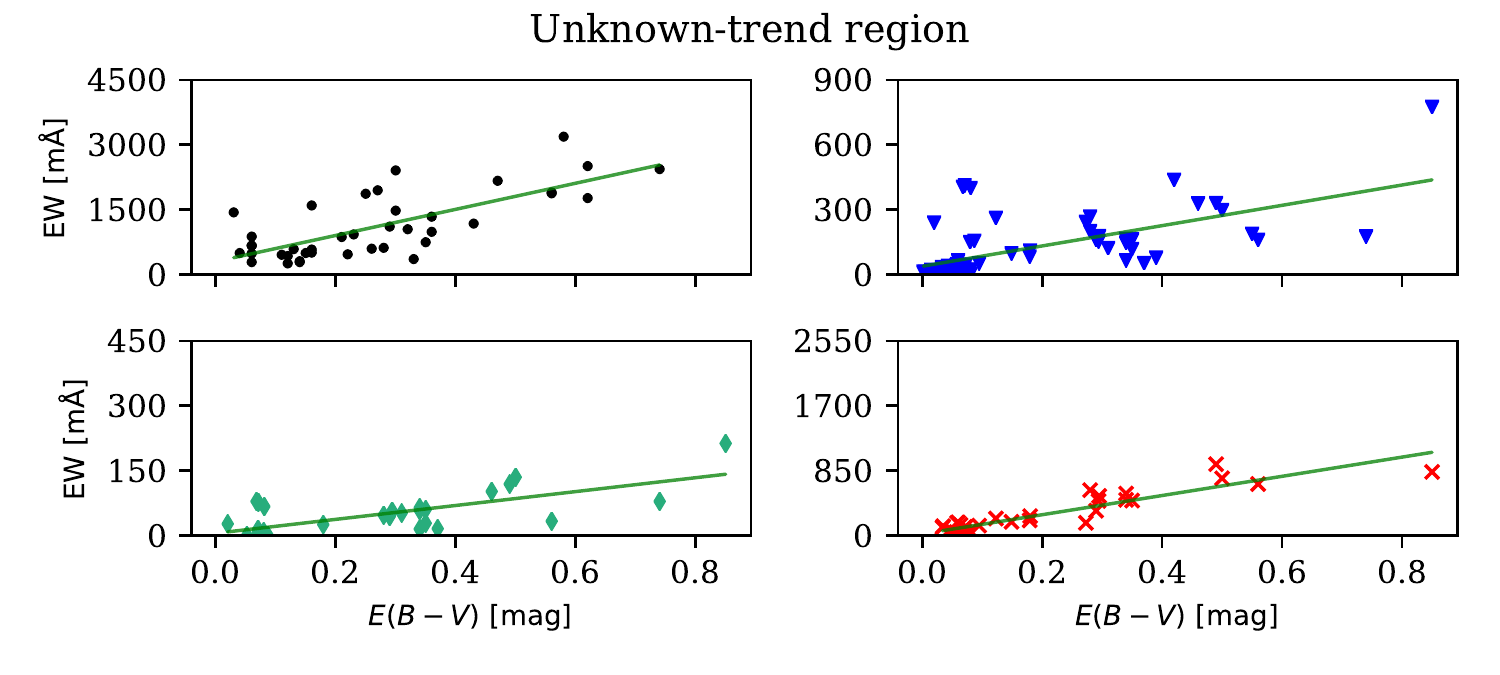}}
  \caption{Correlations between the strength of the DIBs and the colour excess $E(B-V)$ in the unknown-trend region. Green line represents a linear fit to the data with correlation coefficient $r$ and fit parameters $a_0$ and $a_1$. Black dots show 4430$\angstrom$ DIB ($r=0.743 \,,\, a_0=303 \,,\, a_1=3018$), blue triangles 5780$\angstrom$ DIB ($r=0.628 \,,\, a_0=39 \,,\, a_1=470$), teal diamonds 5797$\angstrom$ DIB ($r=0.718 \,,\, a_0=5 \,,\, a_1=161$) and red crosses 6284$\angstrom$ DIB ($r=0.918 \,,\, a_0=21 \,,\, a_1=1258$).}
  \label{figtest6}
\end{figure*}

\begin{figure*}
  \resizebox{\hsize}{!}{\includegraphics{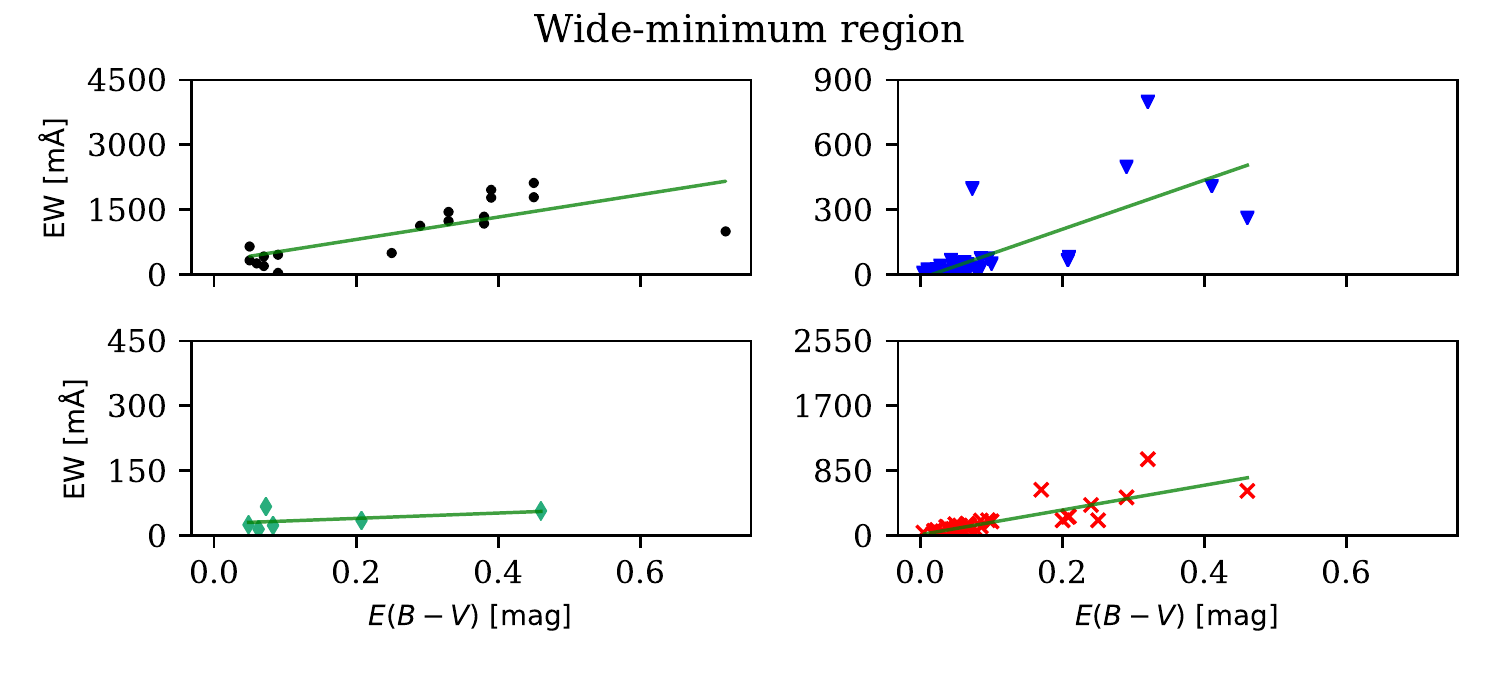}}
  \caption{Correlations between the strength of the DIBs and the colour excess $E(B-V)$ in the wide-minimum region. Green line represents a linear fit to the data with correlation coefficient $r$ and fit parameters $a_0$ and $a_1$. Black dots show 4430$\angstrom$ DIB ($r=0.746 \,,\, a_0=296 \,,\, a_1=2589$), blue triangles 5780$\angstrom$ DIB ($r=0.745 \,,\, a_0=-17 \,,\, a_1=1137$), teal diamonds 5797$\angstrom$ DIB ($r=0.477 \,,\, a_0=27 \,,\, a_1=62$) and red crosses 6284$\angstrom$ DIB ($r=0.821 \,,\, a_0=11 \,,\, a_1=1621$).}
  \label{figtest7}
\end{figure*}

\begin{figure*}
  \resizebox{\hsize}{!}{\includegraphics{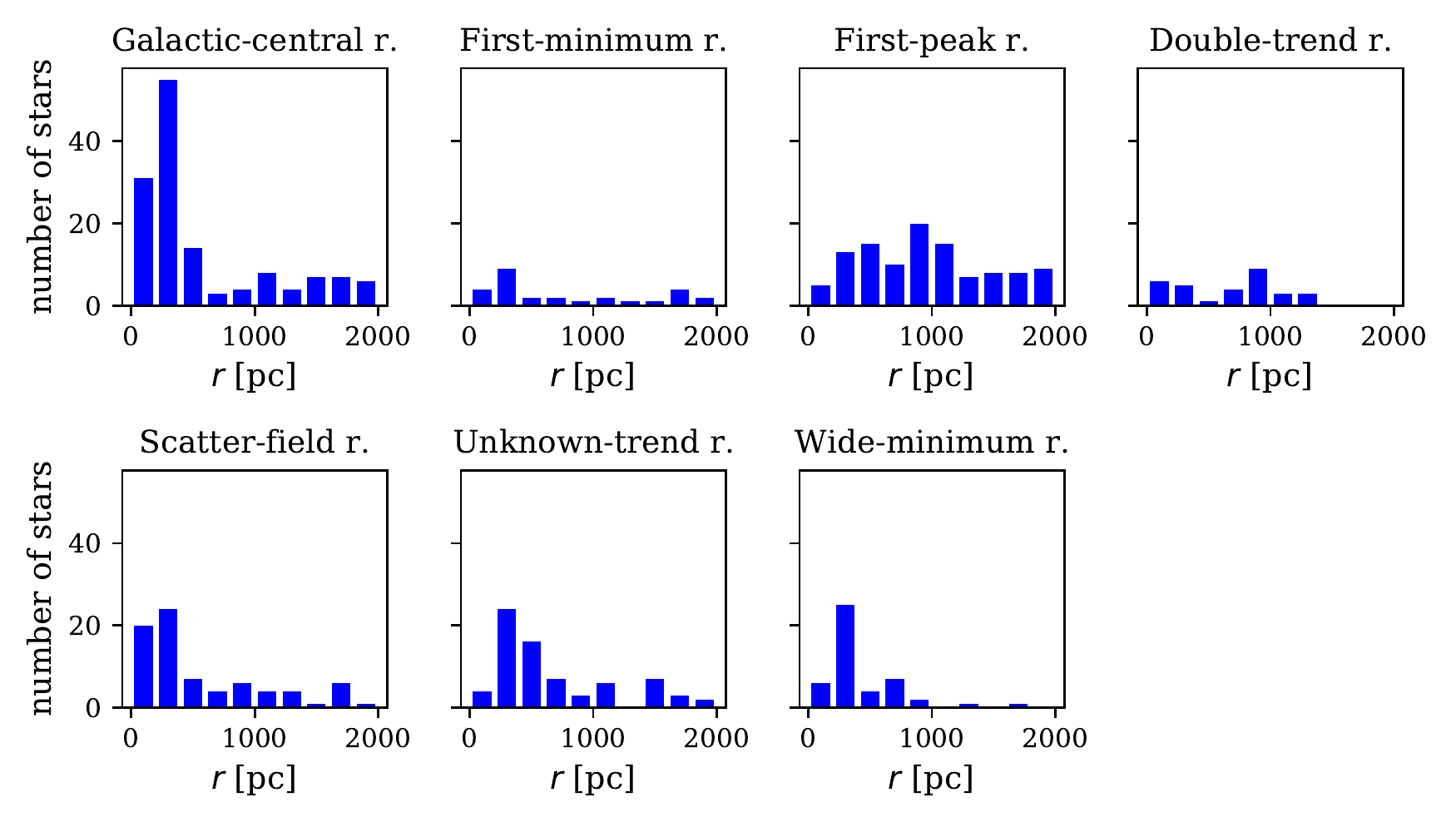}}
  \caption{Distribution of distances (up to 2 kpc) in all regions defined in Table \ref{table:2}.}
  \label{figH1}
\end{figure*}

\begin{figure*}
  \resizebox{\hsize}{!}{\includegraphics{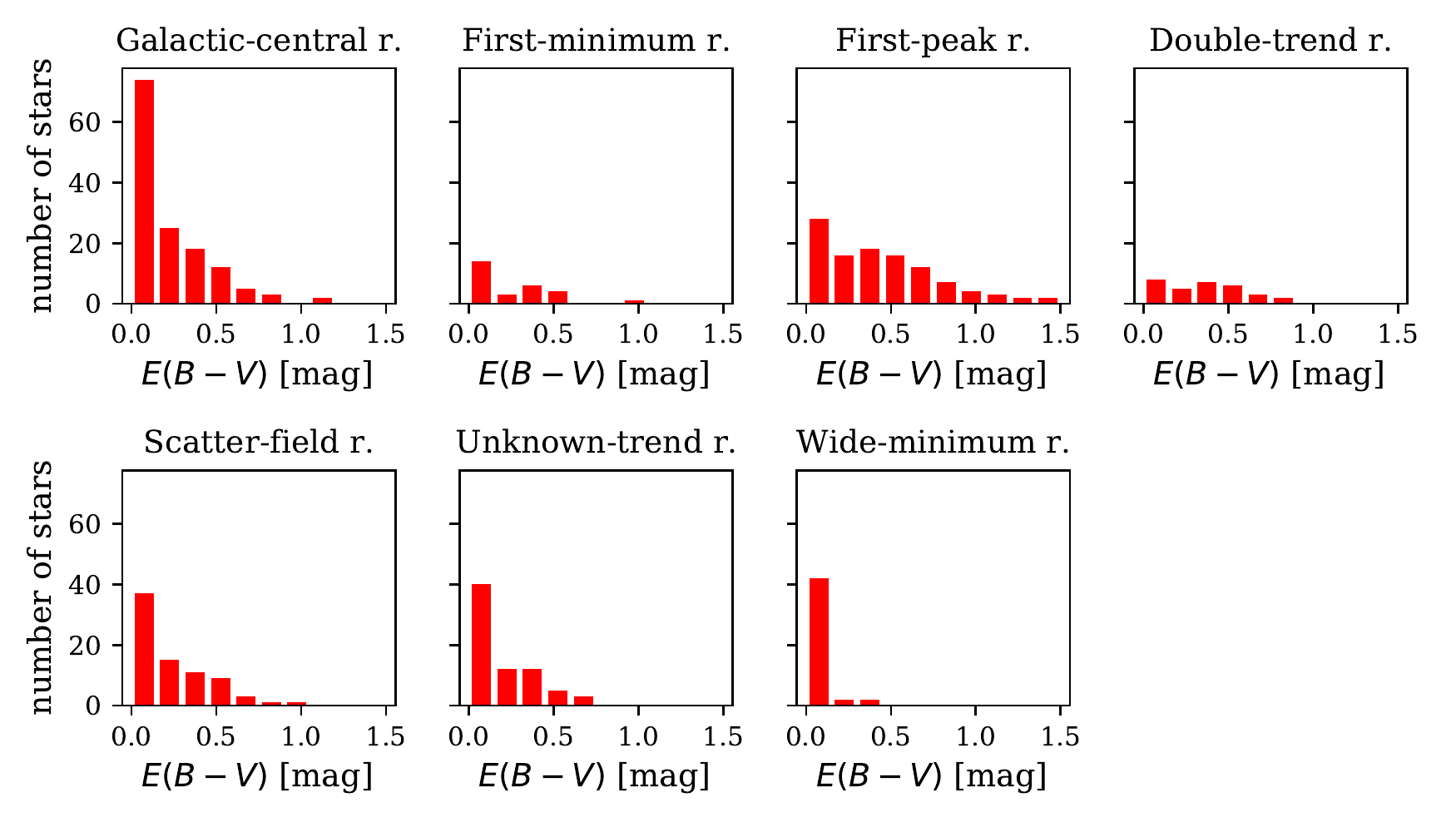}}
  \caption{Distribution of colour excess values (for distances up to 2 kpc) in all regions defined in Table \ref{table:2}.}
  \label{figH2}
\end{figure*}


\bsp	
\label{lastpage}
\end{document}